\newcommand{\arxiv}[1]{}

\newcommand{\pepm}[1]{}

\newcommand{\now}[1]{#1}
\newcommand{\notnow}[1]{}

\arxiv{}

\pepm{}

\now{
\documentclass[PGL,biber]{nowfnt}
\usepackage[utf8]{inputenc}
\renewcommand{\cite}[1]{\citep{#1}} %
}

\usepackage{alltt} %
\usepackage{hyperref}

\usepackage{xspace}

\usepackage[normalem]{ulem} %

\newcommand{\Hex}[1]{\hspace{#1ex}}
\newcommand{\Vex}[1]{\vspace{#1ex}}

\newcommand{\mysec}[1]{\notnow{}\now{\chapter{#1}}}
\newcommand{\mypar}[1]{\notnow{}\now{\section{#1}}}
\newcommand{\mysubpar}[1]{\paragraph{ \Hex{-.75}\rm\uline{#1.}}}
\newcommand{\myemph}[1]{\uline{#1.}~}
\newcommand{\defn}[1]{\uline{#1}}
\newcommand{\term}[1]{\uline{#1}}
\newcommand{\hlit}[1]{\uline{#1}} %

\newcommand{\mathify}[1]{\ifmmode{\mbox{$#1$}}\else\mbox{$#1$}\fi}
\newcommand\m[1]{\mathify{#1}} %
\newcommand\p[1]{\mathify{#1}} %

\newcommand\co[1]{\mbox{\small\tt #1}} %
\newcommand\kw[1]{{\bf #1}} %

\newcommand{\Inc}{{\it Inc}}
\newcommand{\PE}{{\it PE}}
\newcommand{\interp}{{\it interp}}

\newcommand\U{\m{\cup}}
\newcommand\pluseq{~+\!:=\,}
\newcommand\unioneq{~\U\!:=}

\newcommand\IF{\kw{if~}\xspace}
\newcommand\THEN{\kw{~then~}\xspace}
\newcommand\LET{\kw{~let~}\xspace}
\newcommand\IN{\kw{~in~}\xspace}

\newcommand\longto{~~~\Longrightarrow~~~}
\renewcommand\to{\m{\rightarrow}}

\renewcommand\O[1]{\mathify{O(#1)}\xspace}
\newcommand\s[1]{\mathify{\#}\mbox{\co{#1}}} %
\newcommand\vertex{\m{\it vert}\xspace}
\newcommand\indeg{\m{\it indeg}\xspace}
\newcommand\state{\m{\it state}\xspace}
\newcommand\lbl{\m{\it label}\xspace}

\usepackage{enumitem}
\setlist[description]{itemindent=-2ex}

\notnow{}

\title{Incremental Computation:\linebreak[1] What Is the Essence?\pepm{}}

\arxiv{}

\pepm{} %

\now{
\maintitleauthorlist{Yanhong A. Liu\\
Computer Science Department\\ Stony Brook University\\ Stony Brook, NY 11794,
USA\\
liu@cs.stonybrook.edu}

\issuesetup
{%
 copyrightowner={Y.~A.~Liu},
 volume        = xx,
 issue         = xx,
 pubyear       = 2025,
 isbn          = xxx-x-xxxxx-xxx-x,
 eisbn         = xxx-x-xxxxx-xxx-x,
 doi           = 10.1561/XXXXXXXXX,
 firstpage     = 1, %
 lastpage      = 75?
 }

\def\bibdir{../../bib}   %
\addbibresource{\bibdir/strings.bib}
\addbibresource{\bibdir/liu.bib}
\addbibresource{\bibdir/IC.bib}
\addbibresource{\bibdir/PE.bib}
\addbibresource{\bibdir/PT.bib}
\addbibresource{\bibdir/PA.bib}
\addbibresource{\bibdir/AG.bib}
\addbibresource{\bibdir/Lang.bib}
\addbibresource{\bibdir/Algo.bib}
\addbibresource{\bibdir/DB.bib}
\addbibresource{\bibdir/SE.bib}
\addbibresource{\bibdir/Math.bib}
\addbibresource{\bibdir/Sys.bib}
\addbibresource{\bibdir/misc.bib}
\addbibresource{\bibdir/crossref.bib}

\usepackage{mwe}

\author{Liu,Yanhong A.}

\affil{Computer Science Department, Stony Brook University, Stony Brook,
  NY 11794, USA; liu@cs.stonybrook.edu}

\articledatabox{\nowfntstandardcitation}
} %

\notnow{}
\arxiv{}

\now{
\begin{document}
\makeabstracttitle
}

\begin{abstract}
Incremental computation aims to compute more efficiently on changed input
by reusing previously computed results.
We give a high-level overview of works on incremental computation,
and highlight the essence underlying all of them, which we call
incrementalization---the discrete counterpart of differentiation in
calculus.
We present the gist of a systematic method for incrementalization,
and a systematic method centered around it---called Iterate-Incrementalize-Implement---for program design and optimization, as well as algorithm design and
optimization.
We illustrate the methods with example applications in arithmetic
computations, recursive functions, graph analysis, and distributed
algorithms.
At a meta-level,
with historical contexts and for future directions,
we stress the power of high-level data, control, and module abstractions
in developing new and better algorithms and programs as well as their precise
complexities.
\end{abstract}

\pepm{} %

\mysec{Introduction}

As the real world changes continually, computer programs that handle input
from the real world must handle continually changing input.
Furthermore, because any algorithm that solves a nontrivial problem must
proceed in an iterative or recursive fashion, computations 
must be performed on repeatedly changed state.
The general area of \term{incremental computation} studies how to
efficiently handle continually changing input by storing and reusing
previously computed results.
The problem is fundamental and ubiquitous, from graph algorithms to
database queries, and from program analysis to neural network training,
including everything nontrivial.

There was already a large literature on incremental computation even over 30
years ago~\cite{RamRep93}, and the area has grown significantly since
then.\footnote{A Google Scholar search of ``incremental computation'',
  including the quotes, performed on Oct.\ 4, 2025 returned ``About 495 results''
  for the period up until 1992, which includes the period 
  covered by~\cite{RamRep93}, and returned ``About 9,000 results''
  for the period up until the present.} %
To grasp the depth and breadth of the area, it is important to understand
the essence underlying different works and results.

This writing gives a high-level overview of works on incremental
computation---organizing them into \term{three categories}: incremental
algorithms, incremental program-evaluation frameworks, and
incremental-algorithm-and-program
derivation methods---and highlights the essence underlying all of them,
which we call \defn{incrementalization}.
\begin{quote}
Given a
program \m{f}
and an input change operation \m{\oplus}, incrementalization aims to obtain an
incremental version of \m{f} under \m{\oplus}, denoted \m{f'}, that computes on
the changed input more efficiently by reusing results computed before the
change.
\end{quote}
It is the discrete counterpart of \term{differentiation} in calculus.

We then present the gist of a systematic method for
incrementalization~\cite{Liu00IncEff-HOSC}, and a systematic method
centered around it, called \term{Iterate-Incrementalize-Implement (III)},
for program design and optimization, as well as algorithm design and
optimization~\cite{Liu13book}.
\begin{itemize}

\item Systematic incrementalization is a general transformational
  method for deriving incremental programs
  that compute more efficiently by exploiting the \hlit{previous result}, %
  \hlit{intermediate results}, and %
  \hlit{auxiliary information}.
  The method consists of systematic program analysis and transformations
  that are modular, 
  and that are drastically easier and more powerful for programs that use
  high-level
  abstractions.

\item III is a general transformational method for designing and optimizing
  programs, for programs written using different language features---loops
  and arrays,
  set expressions,
  recursive functions,
  logic rules,
  and objects and classes.
  The method is particularly powerful for high-level \term{data
    abstractions} using \term{sets}, as in database programming;
  \term{control abstractions} using \term{recursion}, as in functional
  programming; both data and control abstractions
  using \term{rules}, as in logic
  programming; and \term{module abstraction} using \term{objects}, as in
  object-oriented programming.

\end{itemize}
We will also see that incremental computation is closely related to and
intertwined with \term{partial evaluation}, %
and that optimization by
incrementalization corresponds to \term{integration by differentiation}.

We illustrate the methods with example applications: arithmetic
computations for tabulating polynomials, in relating incrementalization to
differentiation; recursive functions for computing the Fibonacci function,
in relating optimization by incrementalization to integration by
differentiation; graph analysis for computing the transitive closure and
for regular path queries; and distributed algorithms for solving
distributed mutual exclusion.

At a meta-level,
with historical contexts and for future directions,
we stress the power of high-level data, control, and module abstractions in
enabling systematic analysis and transformations for incrementalization,
leading to new and better algorithms and programs as well as their precise
complexities.

At the same time, systematic design and optimization using
incrementalization in turn have helped raise the level of abstractions,
enabling the design of a high-level language for distributed
algorithms~\cite{Liu+12DistPL-OOPSLA,Liu+17DistPL-TOPLAS} and subsequently
a simple unified semantics for logic rules with unrestricted negation,
quantification, and
aggregation~\cite{LiuSto20Founded-JLC,LiuSto22RuleAgg-JLC}.

For future work, significant further effort %
is needed to put both high-level
abstractions and powerful transformation methods into practice, for
developing algorithms and generating programs with both correctness and
performance guarantees.

This writing is an extended version
of an invited article~\cite{Liu24IncEssence-PEPM}.
The main differences include (1) significantly expanded discussions
relating incrementalization with differentiation
(Section~\ref{sec-inc-diff}), and relating optimization by
incrementalization with integration by differentiation
(Section~\ref{sec-inc-opt}), with expanded simple examples in arithmetic
computations (Section~\ref{sec-inc-diff}) and recursion functions
(Section~\ref{sec-inc-opt}), (2) newly added example applications in graph
analysis (Sections~\ref{sec-tc} and~\ref{sec-rpq}) and in distributed
algorithms (Section~\ref{sec-lamutex}), and (3) added discussions about
related concepts throughout, e.g., semi-naive evaluation and magic-set
transformations in Section~\ref{sec-rules} and objects being merely a
change of perspective and live objects being concurrent and distributed
processes in Section~\ref{sec-obj}.

When describing the personal experience of the author, we will write in
the first person. %

\mysec{Incremental computation: three %
categories of studies}
\label{sec-3cat}

Despite the vast amount and variety, %
we organize work on incremental computation into three 
categories.
\begin{description}

\item [Incremental algorithms.]
  Algorithms for computing particular functions, such as shortest paths
  in a graph, under particular kinds of input changes, such as adding and
  deleting graph edges.

  This includes algorithms known as dynamic
  algorithms, online algorithms, and other variants.

\item [Incremental program-evaluation frameworks.]
  Frameworks for evaluating general classes of programs expressed in the
  framework and handling input changes.

  This includes frameworks known as memoization,
  caching, tabling, change
  propagation, and other variants.

\item [Incremental-algorithm-and-program derivation methods.]~\linebreak
  Methods for deriving algorithms and programs that handle input changes
  from given algorithms or programs and given kinds of input changes.

  This includes methods known as finite differencing, maintaining and
  strengthening loop invariants, automatic differentiation, incrementalization,
  and other variants.
\end{description}
This categorization was first developed in 1991 from my %
year-long study of
work in incremental computation. It enabled me %
to formulate systematic
incrementalization~\cite{LiuTei93Drv-TR,LiuTei95Inc-SCP} for my %
Ph.D.\ thesis proposal in May 1992.
Now, more than 
30 years later, after an extensive\notnow{}
further study of the literature, this categorization is further confirmed.

Thanks to technological advances and extensive work by many people, this
new study of the literature has been blessed with tremendous new resources
not available over 30 years ago: the Web, Google Scholar,
free access to earlier literature from major publishers like ACM and
Elsevier,
and to newly available
literature from resources like arXiv and the Computer History Museum.
Given the vast literature spreading in every dimension,
this study has tried to include the earliest sources found, a range of
examples, and latest overviews.

\mypar{Incremental algorithms}
\label{sec-3cat-1}

Algorithms are at the core of computation, aiming to compute desired output
efficiently from given input.
Given certain ways that input can change, an \defn{incremental algorithm}
aims to compute desired output efficiently by maintaining and reusing
previously computed results.
For example, an algorithm for sorting computes sorted output
from the given input, whereas
an incremental algorithm for sorting when an element is
added to or deleted from the input computes sorted output by storing and 
updating the previously sorted output. %

\myemph{Example algorithms} %
Incremental algorithms have been studied since at least the
1960s~\cite{RamRep93}, e.g., for maintaining the shortest distances in a
graph when the length of an edge is increased or
decreased~\cite{murchland1967effect}.  
The area has grown to include a vast number of algorithms for a 
wide variety of problems and variants, 
e.g., incremental parsing~\cite{Ghezzi:Mandrioli:79,wagner1998efficient},
incremental attribute
evaluation~\cite{Reps:Teitelbaum:Demers:83,Yeh:Kastens:88}, incremental
circuit evaluation~\cite{Alpern:Hoover:Rosen:Sweeney:Zadeck:90},
incremental constraint
solving~\cite{Freeman:Maloney:Borning:90,dantam2018incremental}, as well as
many incremental graph problems and
more~\cite{sharp2007incremental,mcgregor2014graph,fan2017incremental,hanauer2022recent}, to reference a few.
Although efforts in this category are directed towards particular
incremental algorithms, an algorithm may apply to a broad class of problems, 
e.g., any attribute grammar, any %
graph, etc. %

Many incremental algorithms, especially those on graphs, are also known as
\defn{dynamic algorithms}, which are called fully dynamic, incremental, or
decremental if the algorithm handles both additions and deletions of edges,
only additions, or only deletions,
respectively~\cite{demetrescu2009dynamic,hanauer2022recent}.  Note that we
use the term ``incremental dynamic algorithms'' to differentiate from uses of
``incremental algorithms'' for the general category.

A class of incremental algorithms is known as \defn{online algorithms},
which process input piece-by-piece in the order that the input is fed to
the algorithm~\cite{borodin2005online,sharp2007incremental}, essentially
corresponding to incremental dynamic algorithms.  For example, an online
algorithm for sorting processes each next element in the input as it comes.
Another class of incremental algorithms is known as \defn{streaming
  algorithms}, which process a sequential stream as input but with limited
memory~\cite{muthukrishnan2005data,mcgregor2014graph},
and thus can only examine the input in a few passes, typically one pass,
and often produce approximate output using what is called a sketch.
For example, a streaming algorithm can sort, say, the 10 smallest elements
but not the entire input.
There are also semi-streaming algorithms that consider a relaxation of the
memory limitation for graph problems~\cite{feigenbaum2005graph}.

A wide range of \defn{efficiency and other measures}, with trade-offs, and
complexity models,
are used for characterizing incremental algorithms.
General incremental algorithms and dynamic algorithms mainly aim to minimize
the times to maintain needed information and return desired
output.
Online algorithms also aim to be competitive in performance against the
case that the entire input is given at the start.
Streaming algorithms aim to use small space and consider also approximation
ratio when output is approximated, for example, for counting the number of
distinct elements when they do not fit in memory.

\mypar{Incremental program-evaluation frameworks}
\label{sec-3cat-2}

Rather than manually developing incremental algorithms for each particular
problem, an \defn{incremental program-evaluation framework} allows
non-incremental programs that are expressed in the framework to run
directly and automatically handles changes supported by the framework to
achieve incremental computation.  For example, a framework can support
caching
the return values of calls to certain functions 
and automatically reuse the cached results when such functions are called again
on the same arguments.

\myemph{Example frameworks} %
Incremental computation appeared in the 1960s as a framework 
using the LISP language for computing function applications as partial
information about the 
input is provided~\cite{Lombardi:Raphael:64,Lombardi:67}. %
``Memo'' functions and machine learning appeared in Nature April 1968 as a
framework for caching and reusing the results of functions~\cite{Michie68}.
Numerous frameworks have since been proposed, e.g.,
\begin{itemize}
  \setlength\itemsep{0ex}
\item 
spreadsheets since the earliest, in 1969, with forward referencing and
natural order recalculation~\cite{pardo1983process,higgins2009grid},
\item 
incremental attribute evaluation frameworks performing change propagation~\cite{Reps:Teitelbaum:Demers:83},
\item 
function caching with improved cache replacement~\cite{Pugh:Teitelbaum:89},
\item 
the INC language with change detailing network~\cite{Yellin:Strom:91},
\item 
incremental reduction in lambda calculus~\cite{Field:Teitelbaum:90,abadi1996analysis},
\item 
logic rule engines with tabling~\cite{TamSat86,CheWar96} and
incremental tabling~\cite{SahaRam03},
\item 
combining change propagation and memoization using dynamic dependency
graphs~\cite{Hoover:92,Acar09},
\item 
combining them also with demand~\cite{Hoover:92,hammer2014adapton},
\item 
and incremental graph processing for vertex-centric computations~\cite{gong2024ingress},
\end{itemize}
to mention a few.

\defn{Caching}---also known as \defn{memoization} and 
\defn{tabling}%
---is the key idea in all these
frameworks to enable reuse of previously computed results.  On top of it,
with previously computed results saved, \defn{change propagation} aims to
compute only results that depend on the changes in input.  Additional
improvements to these techniques and combinations of them,
e.g.,~\cite{Hoover:92,AcaBleHar03,Acar09,hammer2014adapton,gong2024ingress}, can
enable more refined reuse in more specialized cases,
all by expressing the given problem using the mechanisms supported by the
framework,
without
manually writing a particular incremental program for each particular
problem.

At the same time, %
when such frameworks are used, no explicit
incremental version of an application program is derived 
and run by a standard evaluator.  Also, any input change to an
application program is captured in a form that the framework can handle,
which is limited for each framework.  As a result, these solutions to the
incremental computation problem for particular application functions and
changes are not readily comparable with explicitly developed incremental
algorithms and programs for those functions and changes.

\mypar{Incremental-algorithm-\pepm{}and\pepm{}-program derivation methods}
\label{sec-3cat-3}

Instead of developing incremental algorithms for each problem in an ad hoc
way or developing an incremental evaluation framework for problems expressed in that
framework, an \defn{incremental-algorithm-and-program derivation method} aims to
derive explicit
incremental algorithms and programs from non-incremental programs and given
input change operations using semantics-preserving program transformations.
For example, a method can start with a sorting program and an input change
operation that adds a new input element and derive an algorithm and program
that inserts the new element into the previously sorted result.

\myemph{Example methods}
Early proponents of high-level languages supporting recursive
functions~\cite{mccarthy1959lisp,burstall1968pop} and
sets~\cite{schwartz1971set,earley1974high} pioneered the study of efficient
language
implementations~\cite{mccarthy1960recursive,%
cocke1970programming,%
earley1971toward,%
Darlington:Burstall:73,%
schwartz1974automatic%
},
especially %
transformations to avoid repeated expensive computations in
recursion~\cite{%
mccarthy1962efficient,%
Schwartz:75,%
Burstall:Darlington:77%
}
and in iteration~\cite{Ear76,FonUll76,Paige:Schwartz:77}.
Significant effort has been devoted to such %
transformation methods for incremental computation, e.g., 
\begin{itemize}
  \setlength\itemsep{0ex}
\item 
transformation techniques for 
tabulation~\cite{mccarthy1962efficient,Schwartz:75,Bird:80}, 
\item
maintaining and strengthening loop invariants~\cite{Dijkstra76,Gries:81,Gri84},
\item
iterator inversion for converting from a batch to an incremental
algorithm in VERS2~\cite{Ear76},
\item
finite differencing of set expressions in
SETL~\cite{Paige:Schwartz:77,Pai81thesis,PaiKoe82},
\item
deriving incremental view maintenance in 
databases~\cite{KoePai81,Pai84,HorTei86,CerWid91,koch2010incremental,abeysinghe2022efficient},
\item
promotion and accumulation~\cite{Bird:84:b,Bird:85},
\item
finite differencing for functional programs in KIDS~\cite{Smith:90,Smith:91},
\item
incrementalization for recursive functions~\cite{LiuTei95Inc-SCP,Liu+98Scic-TOPLAS,Liu+01SIEC-SCP,Liu00IncEff-HOSC},
\item
generating incremental object queries~\cite{Liu+05OptOOP-OOPSLA,RotLiu08OSQ-GPCE,Liu+16IncOQ-PPDP},
\item
static differentiation for 
lambda calculus~\cite{cai2014theory,giarrusso2019incremental},
\item
incrementalizing graph algorithms~\cite{fan2021incrementalizing},
\end{itemize}
and more, including automatic
differentiation~\cite{Rall81} that is particularly
important for machine learning~\cite{baydin2018automatic}.

\defn{Transformations for recursive functions},
e.g.,~\cite{Burstall:Darlington:77,Bird:80,Bird:84:b,Smith:91,Liu00IncEff-HOSC},
can be general just as recursive functions are, but are also limited by the
structure of recursion.
The generality can be seen from the wide variety of transformations
on recursive
functions, e.g.,~\cite{Feather86survey,Pettorossi:Proietti:96}.
The limitation is especially notable on list, the main data type used in
recursive functions, because list elements must be traversed from head to
tail in a linear order, which makes it exceedingly complex and inefficient to
access an element in the middle of the list,
not to mention more complex nested lists.
Overall, it is challenging to develop systematic and automatic
transformations that are general~\cite{Smith:90,Blaine:Goldberg:91,Liu95CACHET-KBSE}.

\defn{Transformations for set queries},
e.g.,~\cite{Ear76,FonUll76,Paige:Schwartz:77,PaiKoe82,Pai84,CerWid91,Liu+06ImplCRBAC-PEPM,RotLiu08OSQ-GPCE,koch2010incremental,Liu+16IncOQ-PPDP}, %
can be systematic and automatic, but unlike recursive functions, these queries are
not general, i.e., Turing-complete.
Basically, fixed but powerful rules, and even meta methods, can be
developed for transforming each kind of high-level operation on sets into
more efficient incremental operations when the sets are updated.
In particular, Paige's finite differencing of set
expressions~\cite{Pai81thesis,PaiKoe82} led to new and faster
algorithms for many challenging problems, e.g., 
\begin{itemize}
  \setlength\itemsep{0ex}
\item 
attribute closure~\cite{Paige:Henglein:87} in database design, 
\item 
partition refinement~\cite{Paige:Tarjan:87} for model checking, 
\item 
tree pattern matching~\cite{Cai:Paige:Tarjan:92} for program transformation, 
\item 
DFA minimization~\cite{Keller:Paige:95} 
and regular expression to DFA conversion~\cite{Chang:Paige:97} in automata theory, 
\item
copy elimination~\cite{GoyPai98} for compiler optimization, 
\end{itemize}
and more. %

Overall, these transformation methods, starting from pioneering works such
as~\cite{mccarthy1962efficient,Schwartz:75,Dijkstra76}, help achieve the
general goal of correctness-by-construction---starting from a higher-level
program or specification that is inefficient or even not directly
executable, and deriving a more efficient, incremental program.

\mysec{Essence of incremental computation}
\label{sec-essence}

The three categories of incremental computation are clearly different
from each other---particular algorithms vs.\ general program-evaluation
frameworks vs.\ derivation methods for algorithms and programs.  What is the
essence of incremental computation in all of them?
Additionally, how is incremental computation related to partial evaluation?

In fact, it is understanding this essence and following the
transformational approach for partial evaluation that enabled the
development of the systematic
method for incrementalization discussed in Section~\ref{sec-inc}.

\mypar{Incrementalization}

We first define incremental programs before discussing incrementalization as
the essence of incremental computation.

Given a program $f$ and an operation $\oplus$, a program $f'$ is called an
\defn{incremental version} of $f$ under $\oplus$ if $f'$ computes
$f(x\oplus y)$ efficiently by using $f(x)$.  Precisely,
\begin{equation}
  \label{eqn-s1}
  f(x) = r  \longto f'(x,y,r) = f(x\oplus y)
\end{equation}
That is, if the result of \p{f(x)} is \p{r}, then \p{f'} can use \p{x},
\p{y}, and \p{r} in computing the result of $f(x\oplus y)$.

Note that \p{f} and \p{\oplus} are just two functions. An input \p{x} to
\p{f} can have any structure, e.g., a tuple \p{(x_1,...,x_k)}.  So can a
value of parameter \p{y}. Operation \p{\oplus} can be any function that
takes an old input \p{x} to \p{f} and a value of parameter \p{y} and
returns a new input $x\oplus y$ to \p{f}.
Note also that just as \m{x \oplus y} captures how the input changes from
the previous input \m{x}, \m{f'(x,y,r)} captures how the output changes
from the previous output \m{r}.

Given a program \p{f} and an operation \p{\oplus},
\defn{incrementalization} is the problem of finding an incremental version
\p{f'} of \p{f} under \p{\oplus}.

Incrementalization is the essence of incremental computation.  To see this,
we examine how each of the three categories in Section~\ref{sec-3cat}
achieves incrementalization.
\begin{itemize}

\item \myemph{Incremental algorithms}
It is easy to see that an explicit incremental algorithm, 
as discussed in Section~\ref{sec-3cat-1},
corresponds to an
incremental version \p{f_0'} of a particular function \p{f_0} under a
particular kind of input change operation \p{\oplus_0}:
\begin{equation}
  f_0(x) = r  \longto f_0'(x,y,r) = f_0(x\oplus_0 y)
\end{equation}
For examples, \p{f_0} takes an input list \p{x} and computes a sorted list
\p{r}, \p{\oplus} takes a list \p{x} and a new element \p{y} and returns a
new list with element \p{y} added to \p{x }, and \p{f_0'} inserts \p{y}
into \p{r} in the right place instead of sorting the new list from scratch.

Note that function \p{f_0} and thus \p{f_0'} may return values that have
any structure, e.g., a map mapping each pair \p{(u,v)} of vertices in a
graph to the shortest distance from \p{u} to \p{v}.  In general, function
\p{f_0'} may be computed in two steps: (1) incrementally maintain
appropriate values, including possibly additional values, when the input is
changed, and (2) retrieve or query, from the maintained values, a particular
return value when the value is used.

Also, operation \p{\oplus} may express different kinds of input changes,
e.g., adding or deleting an edge \m{(u,v)} to a set \p{e} of graph
edges---as for fully dynamic graph
algorithms~\cite{demetrescu2009dynamic,hanauer2022recent}---depending on the
value of a parameter, say, \p{tag}. 
That is, \m{\oplus} can take \m{e}
and \m{(u,v,tag)} and return \m{e \cup \{(u,v)\}} if \m{tag=\co{'add'}}
and \m{e-\{(u,v)\}} if \m{tag=\co{'del'}}.

\item \myemph{Incremental program-evaluation frameworks}
An incremental evaluation framework,
as discussed in Section~\ref{sec-3cat-2},
corresponds to an incremental version
\p{eval'} of an evaluator \p{eval} under an input change operation
\p{\oplus}, where the input \p{x} to \p{eval} is a pair: a function \p{f}
and an input \p{data} to \p{f}.  Precisely,
\begin{equation}
\begin{array}{c}
eval(f,data) = r \longto \pepm{}\now{\\}
eval'((f,data),y,r) = eval((f,data)\oplus y)
\end{array}
\end{equation}
That is, \p{eval'} is in itself an incremental algorithm and program,
handling a certain kind of change to its input.
For example, a function caching framework can incrementally execute any
program written with
caching directives, by capturing all changes as changed functions and
arguments in function calls, caching results of certain calls, and reusing
cached results for calls encountered again with same functions and arguments.

For another example, an incremental attribute evaluation
framework~\cite{Reps:Teitelbaum:Demers:83} can solve any tree analysis
problem specified using supported attribute equations, by capturing all
changes to the input tree as subtree replacements that 
the framework handles and running a specific incremental change-propagation
attribute evaluation algorithm.

Note that an input change operation for \p{eval} can in general change both
\p{f} and \p{data}, e.g., %
in~\cite{Field:Teitelbaum:90,abadi1996analysis,huang2024lightweight},
but most incremental evaluation frameworks handle only changes to \p{data},
e.g.,~\cite{Reps:Teitelbaum:Demers:83,Hoover:92,AcaBleHar03,Acar09,hammer2014adapton,gong2024ingress},
especially frameworks for compiled languages.

\item \myemph{Incremental-algorithm-and-program derivation methods}
An incremental algorithm-and-program derivation method,
as discussed in Section~\ref{sec-3cat-3},
is simply a method
\p{\Inc} for deriving an incremental version \p{f'} of \p{f} under
\p{\oplus}, given \p{f} and \p{\oplus}:
\begin{equation}
  \Inc(f, \oplus) = f'
\end{equation}
where \m{f}, \p{\oplus}, and \p{f'} together satisfy (\ref{eqn-s1}).
For example, finite differencing of set
expressions~\cite{Paige:Schwartz:77,PaiKoe82} can automatically derive
incremental maintenance of complex set expressions under various set update
operations by using a collection of finite differencing rules---each for a kind of
set expression and a few update operations---and using the chain rule to
handle nested expressions.  It has been used for optimizing bodies of loops
in set languages~\cite{Paige:Schwartz:77,PaiKoe82} and also efficient
database view maintenance~\cite{KoePai81,Pai84}. %

Note that \p{f} and \p{\oplus} can be any functions or operations written
in the language that the method applies to.  Higher-level or more limited
languages allow more systematic and automatic derivations, while
lower-level and more general languages make the transformations more
challenging.
\end{itemize}

From these, we see that works in all three categories of incremental computation
aim to obtain incremental algorithms or programs for computing a function
incrementally under an input change operation.  This is exactly the essence
of incremental computation, and is what we call incrementalization.

Each work differs in the particular \p{f} and \p{\oplus} considered, the
particular evaluator and kind of changes considered, and the particular
language in which \p{f} and \p{\oplus} are written.

The challenge in all cases is how to achieve better incrementalization more
systematically for more general problems and languages, and ideally achieve
even the best incrementalization possible fully automatically.

\mypar{Incremental computation vs\pepm{}.\ partial evaluation}

An area closely related to and intertwined with incremental computation is
partial evaluation~\cite{Jon+93PE}. 

\defn{Partial evaluation}, also called \defn{specialization}, considers a
program \p{f} whose input has two parts, in the form of \m{(x_1,x_2)}, 
and aims to
evaluate \p{f} as much as possible on a given value \m{x_1} of the first part,
yielding a partially evaluated program \m{f_{x_1}}:
\begin{equation}
  \PE(f, x_1) = f_{x_1}
\end{equation}
so that when given a value \m{x_2} for the second part, \m{f_{x_1}(x_2)}
can compute the result \p{res} of \m{f(x_1,x_2)} more efficiently.
That is, suppose
\begin{equation}
  f(x_1, x_2)=res
\end{equation}
we have
\begin{equation}
  \label{eqn-PE}
  PE(f, x_1) = f_{x_1}, \quad
  f_{x_1}(x_2) = res
\end{equation}

The existence of some program \m{f_{x_1}} that satisfies~(\ref{eqn-PE}) is
actually established by Kleene's \defn{\m{s}-\m{m}-\m{n}
  theorem}~\cite{soare1999recursively}, with \m{x_1} representing \p{m} of
\m{m+n} arguments of \p{f}, and \m{x_2} representing the remaining \p{n}
arguments of \p{f}, even though the proof does not say to evaluate \m{f} on
\m{x_1} as much as possible.
The theorem says that, for any function \p{f} of \m{m+n} arguments and any
given values of \m{m} of the arguments, there exists a function of \m{n}
arguments that, on any given values of the \m{n} arguments, computes \p{f}
on these given values of the \m{m+n} arguments.

Partial evaluation is especially important and interesting in \linebreak\term{connecting
compilers with interpreters}.  Consider an interpreter \p{\interp} that takes
program \p{prog} and data \p{data} as input and returns result \p{res} as
output, i.e.,
\begin{equation}
  \label{eqn-interp}
  interp(prog, data)=res
\end{equation}
We can see the following:
\begin{enumerate}
\item Applying \p{\PE} to \p{\interp} and \p{prog} in (\ref{eqn-interp}),
  we obtain
\begin{equation}
  \label{eqn-PE_interp}
  \PE(\interp, prog) = \interp_{prog}, \quad
  \interp_{prog}(data) = res
\end{equation}
  That is, \p{\interp_{prog}} is like a compiled program for \m{prog},
  because it takes only \p{data} and returns the result \p{res}.

\item Applying \p{\PE} to \p{\PE} and \p{\interp} in (\ref{eqn-PE_interp}),
  we obtain
\begin{equation}
  \label{eqn-PE_PE}
  \PE(\PE,\interp) = \PE_{\interp}, \quad
  \PE_{\interp}(prog) = \interp_{prog}
\end{equation}
  That is, \m{\PE_{\interp}} is like a compiler, because
  it takes \p{prog} and returns a compiled program \m{\interp_{prog}}.

\item Applying \p{\PE} to \p{\PE} and \p{\PE} in (\ref{eqn-PE_PE}),
  we obtain
\begin{equation}
  \PE(\PE,\PE) = \PE_{\PE}, \quad
  \PE_{\PE}(\interp) = \PE_{\interp}
\end{equation}
  That is, \p{\PE_{\PE}} is like a compiler generator, because
  it takes \p{\interp} and returns a compiler \m{\PE_{\interp}}.
\end{enumerate}

Interestingly, incremental computation and partial evaluation were related
physically in 1993, at the 20th ACM SIGPLAN-SIGACT Symposium on Principles
of Programming Languages (POPL), as two invited tutorials with those names,
even with articles in the proceedings~\cite{RamRep93,consel1993tutorial}.
To the best of my knowledge, it was the first time that POPL had any tutorials and many years
before it had any again. %

It is also interesting that the earliest work I %
could find that proposes a
general form of partial evaluation calls it ``incremental computer'' and
``incremental computation''~\cite{Lombardi:Raphael:64,Lombardi:67},
although ``partial evaluation'' is also mentioned 
for one case.
On the other hand, the earliest work I %
could find on a form of incremental
computation in databases by monitoring changes and avoiding recomputation
calls it ``partial evaluation''~\cite{buneman1979efficiently}, and no
``incremental'' or ``propagation'' was mentioned.

There are actually interesting exact relationships between incremental
computation and partial evaluation, going both ways.
\begin{itemize}

\item On the one hand, partial evaluation can be viewed as a special case
  of incremental computation.

  Consider any program \p{f} whose input has two parts \m{(x_1,x_2)}.
  Define \m{\oplus} as:   
  \m{(x_1,x_2) \oplus y = (x_1,y)}
  Then an incremental program will aim to reuse computed values on the
  \m{x_1} part of the input as much as possible and thus compute with the
  new parameter \m{y} as the changed part more efficiently.  This is
  exactly what partial evaluation aims to do.

\item On the other hand, incremental computation can be viewed as a special
  case of \defn{generalized partial
    evaluation}~\cite{Futamura:Nogi:88:1,futamura1991essence}.

  Generalized partial evaluation aims to evaluate as much as possible on
  any given information about program \p{f} and input \p{x}, not limited to \p{x}
  being a given value of one part of the input.
  For a simple example, the information may be \m{x>5}, so any
  computation in \m{f} in a branch with condition, say, \m{x<3} can be
  removed.

  With that, consider the given information \m{x = x_{prev} \oplus y} and
  \m{f(x_{prev}) = r}. Using this information to compute efficiently is
  exactly what incremental computation does.

\end{itemize}
Note that 
generalized partial evaluation essentially aims to optimize any program
to be the most efficient
using any given information.  Whether a program can be made more efficient
is an undecidable problem in general.
However, powerful methods can be developed for special kinds of input
information such as that for partial evaluation and incremental computation.
In fact, systematic incrementalization described in Section~\ref{sec-inc}
uses specialization of \m{f(x)} in specific contexts, %
which can also be
regarded as a special case of generalized partial evaluation.

\mysec{Systematic incrementalization---using previous result,
  intermediate results, and auxiliary
  values}
\now{\chaptermark{Systematic incrementalization}}
\label{sec-inc}

We give a highly distilled overview of a general
and systematic method for incrementalization,
first developed 
in my Ph.D.\ work~\cite{Liu96thesis}. %
The method is general in that it applies to any language for writing \p{f}
and \p{\oplus}.  It is systematic in that it consists of systematic program
analysis and transformations.

Note, however, that higher-level languages allow easier and better analysis
and transformations, and thus better incrementalization, enabling more
drastic optimization by incrementalization discussed in
Section~\ref{sec-opt}.

The transformational approach was inspired by the use of systematic
analysis and transformations in partial evaluation~\cite{Jon+93PE}, and
made possible by focusing on the problem of incrementalization as we
defined, not general program improvement using general unfold-fold
transformations that require ``eureka''~\cite{Burstall:Darlington:77}.

We conclude the section by relating incrementalization to differentiation
and furthermore to integration in calculus.

\mypar{A systematic transformational method for incrementalization}
\label{sec-inc-trans}

The overall method was developed incrementally, by solving three key and
increasingly harder problems, %
giving increasingly greater incrementality.
Each next problem was better understood after the previous problems were
solved, and was also more easily solved by reducing to the previous
problems.

The key idea for solving all three problems is exactly to analyze and
transform \m{f(x\oplus y)}---expanding it and separating computations
that depend on \m{x} from those that depend on \m{y}, and then storing and
reusing values that were computed on \m{x}.

\begin{description}
\item[P1. Exploiting the previous result.]
  The problem is to use the return value \m{r} of \m{f(x)} in computing
  \m{f(x\oplus y)}.  The most straightforward use is: after transforming
  \m{f(x\oplus y)} to separate computations on \m{x} and on \m{y}, if there
  is a computation on \m{x} that is exactly \m{f(x)}, then replace it with
  \m{r}.

  More powerful uses are by \defn{exploiting data structures and control
    structures} in \m{f(x)}:

  If \m{r} is a structured data, e.g., a tuple whose first component is
  \m{f_1(x)}, then 
  a computation \m{f_1(x)} in \m{f(x\oplus y)} can be replaced with a
  retrieval \m{1st(r)}, the first component of \m{r}.

  If \m{f_1(x)} is computed only inside a branch with a condition, e.g.,
  \m{x \neq null}, in \m{f(x)}, then replacement of \m{f_1(x)} in
  \m{f(x\oplus y)} must be in a branch where \m{x \neq null} holds.

\item[P2. Caching intermediate results.] The problem is to use helpful
  intermediate
  values computed while computing \m{f(x)}, not just the return value. But,
  which values and how to use them?  

  With P1 solved, the conceptually simplest solution for P2 is
  \defn{cache-and-prune}: (1) transform \m{f} to cache all intermediate values
  in the return value, yielding \m{f_{cache}}, (2) use 
  P1 to incrementalize \m{f_{cache}}, yielding \m{f_{cache}'}, and (3)
  prune \m{f_{cache}'} and \m{f_{cache}} to retain only values needed
  for obtaining the original return value. %
  Step (1) is straightforward.  Step (3) identifies values in the result
  \m{r_{cache}} of \m{f_{cache}(x)} that are used in computing the
  original return value in \m{f_{cache}'(x,y,r_{cache})}; 
  this process may repeat because computing the identified values in 
  \m{f_{cache}'(x,y,r_{cache})} may use other values in \m{r_{cache}}.

  Alternatively, \defn{selective caching} can (1) use and extend solutions
  to P1 to identify, in
  \m{f(x\oplus y)},
  computations on \m{x}
  that are intermediate computations in \m{f(x)},
  (2) transform \m{f} to cache such computation results in the return
  value, and (3) use P1 to incrementalize the transformed program.
  This process may repeat because computing the cached values after
  \m{\oplus} may use values of other intermediate computations.

\item[P3. Discovering auxiliary information.] The problem is to use also
  values not computed in \m{f(x)} at all but that can help in computing \m{f(x\oplus
    y)}. But, where to find such values, and how to use them?

  With P1 and P2 solved, the simplest solution for P3 is to (A) use and
  extend P1 to identify, in
  \m{f(x\oplus y)},
  computations on \m{x} that are not computed at all in \m{f(x)}, as
  \defn{candidate auxiliary values},
  and (B) use and extend P2 to extend \m{f} to also return the candidate
  auxiliary values.

  This process may repeat, because computing the auxiliary values after
  \m{\oplus} may need other auxiliary values. Also, because these values are
  not computed in the original program, cost analysis is needed to use only
  auxiliary values that help incremental computation.

  This enables the use of a general class of auxiliary information that can
  be found using systematic analysis and transformations.

\end{description}
The resulting overall method is modular.
The method can be fully automated because the analysis and transformations
used can be conservative and fully automatic~\cite{ZhaLiu98AutoInc-ICFP}.

The method is described in more detail in an overview
article~\cite{Liu00IncEff-HOSC} and in the last chapter of a
book~\notnow{}\now{\citep}[Chapter 7]{Liu13book}, with references to
detailed analyses and transformations for
P1~\cite{LiuTei93Drv-TR,LiuTei95Inc-SCP},
P2~\cite{LiuTei95Cir-PEPM,Liu+98Scic-TOPLAS}, 
and P3~\cite{Liu+96Dai-POPL,Liu+01SIEC-SCP}.
The book exists owing to repeated suggestions
and encouragements from Neil Jones, Michel Sintzoff, and others,
and the overview article exists owing to
warm suggestions and detailed comments from Olivier Danvy.

The method is powerful in deriving incremental algorithms and programs,
e.g., different incremental sorting algorithms from batch sorting
programs~\cite{LiuTei95Inc-SCP}, incremental reporting for enterprise
resource planning systems~\cite{nissen2007funsetl}, and
incremental queries in access control, distributed algorithms,
probabilistic inference, etc.~\cite{Liu+16IncOQ-PPDP}.
The method is even more powerful when used in general program
optimizations, as introduced in Section~\ref{sec-inc-opt}.

\mypar{Incrementalization---differentiation in discrete domains}
\label{sec-inc-diff}

We can see that incrementalization corresponds to differentiation in
calculus, except that incrementalization takes place in discrete domains as
opposed to continuous domains.
We even use the same notation \m{f'} for the incremental version of \m{f}
from incrementalization as the derivative of \m{f} from differentiation.

\mysubpar{From changes in input to changes in output}
Both incrementalization and differentiation study changes in the output of
functions given changes in the input.  While differentiation yields the
derivative of a function where input changes are infinitesimal,
incrementalization yields an incremental version of a function under an
input change operation, like a \term{finite difference}~\cite{jordan1965calculus}
when the function is defined only at discrete points.
Precisely, in the definition of incremental version,
\begin{itemize}
  \setlength\itemsep{0ex}
\item \m{x \oplus y} captures exactly how the input changes from the
  previous input \m{x}, and
\item \m{f'(x,y,r)} captures exactly how the output changes from the
  previous output \m{r}.
\end{itemize}
Note that in the incremental version \m{f'(x,y,r)} that computes \m{f(x
  \oplus y)}, if all computations on \m{x} have been replaced with uses of
\m{r}, then \m{x} is unused and can be removed, and \m{f'(y,r)} computes
the new output by using only the old output \m{r} and the input change
\m{y}.

Looking more deeply, there are many further correspondences. The following
are a few primary ones:
\begin{itemize}
\item \myemph{Chain rule}
  To incrementalize a function defined by composing two smaller functions,
  an incremental version of the inner function acts as the input change
  operation of the outer function, enabled by caching the intermediate
  result of the inner function; this corresponds to the chain rule for
  differentiation.
\item \myemph{Discontinuity removal}
  To incrementalize a function that requires more than the previous result
  and intermediate results, discovering auxiliary information corresponds
  to discontinuity removal, where the function definition at a specific
  point is added or modified so that the derivative of the function can
  exist.
\item \myemph{Higher-order derivative}
  Repeating incrementalization with caching selectively and with discovered
  auxiliary values
  corresponds to computing each successive derivative of the previous
  derivative, i.e., computing higher-order derivatives.

\end{itemize}

\mysubpar{Challenges posted by discrete domains}
In general, however, handling changes in discrete domains makes
incrementalization much more difficult than differentiation, because
functions must be continuous for differentiation, but functions on discrete
domains have mostly holes.
This %
loss of smoothness is the main source of difficulty.  
It is also why discrete optimization is generally harder than continuous
optimization and often results in less efficient solutions, e.g., the 0-1
knapsack problem is NP-hard whereas the continuous knapsack problem can be
solved in polynomial time.

More specifically, while in simple cases, incrementalization succeeds by
just exploiting the previous result or intermediate results, in complex
cases, auxiliary information is essential.
\begin{itemize}

\item As described above, discovering auxiliary information corresponds to
  discontinuity removal.  While the latter requires defining function
  values at some specific points, the former requires defining entire
  auxiliary functions 
  over discrete domains.

\item Also as described above, finding such auxiliary information, even
  with our systematic method, may in general require repeated
  incrementalization, i.e., computing higher-order derivatives, but with
  complex cost and trade-off issues to consider.

\end{itemize}
Additional effort is also needed to handle other issues caused by updates
in discrete domains.  The following is a very small example.

Consider the example of computing the square of a number \m{x}.  We know
that
\begin{center}
  if \m{f(x) = x^2}, then its derivative is \m{f'(x) = 2x}  
\end{center}
But if \m{x} can only be integers, even for the smallest change \m{x\oplus
  y = x+1,}\footnote{\m{y} is a dummy unused variable in this example.} and
given \m{f(x)=r}, additional care is needed:
\begin{itemize}

\item \myemph{Exploiting algebraic properties} 
This allows us to separate   out \m{x^2} from \m{(x+1)^2}, yielding
\begin{equation}
\begin{array}{@{}l@{}}
  f(x\oplus y) = f(x+1) = (x+1)^2
                        = x^2 + 2x + 1\\
  \mbox{so}~ f'(x,r) = r+2x+1
\end{array}
\end{equation}
That is, the increment in output is not \m{2x} but \m{2x+1}, because the
change to \m{x} is not infinitesimal.

\item \myemph{Maintaining result at update} 
  Incrementally maintaining the result \m{r} of \m{x^2} at the
  update to \m{x}, i.e., \m{x \pluseq 1}, depends on whether the
  maintenance is done before or after the update:
\begin{equation}
  \label{eqn-x2inc}
\begin{array}{@{}r@{}}
  \mbox{if before \m{x \pluseq 1}, we must do \m{r \pluseq 2x +1}}\\
  \mbox{if after~~~\m{x \pluseq 1}, we must do \m{r \pluseq 2x -1}}
\end{array}
\end{equation}
Note that this issues is not due to imperative programming with updates,
because in any programming, one must decide, when computing the new \m{r}
incrementally using \m{x}, whether the new or old \m{x} is used.
This is analogue to {forward vs.\ backward difference} when computing a
finite difference.

\item \myemph{Maintaining result on demand}  Maintaining the result \m{r} can
  also be done on demand at uses of \m{r} instead of eagerly at updates to
  \m{x}.

  This could save significantly when multiple updates may happen before a
  use and may even cancel each other out, but it can also be costly to
  track many changes and do batch incremental maintenance compared with
  computing from scratch.

  In general, computations in incremental maintenance can be done, in their
  dependency order, at any point after an update and before the result is
  next used.  A careful trade-off analysis is needed to decide when and
  where to maintain what values.

\item \myemph{Considering costs and trade-offs} This is essential for
  deciding what to do in general.  For example, in the maintenance \m{r
    \pluseq 2x + 1} in (\ref{eqn-x2inc}), it is assumed that doubling and
  addition are cheap compared with squaring.  If only addition is cheap,
  \m{2x} could be replaced with \m{x+x}.

  In fact, in either case, \m{2x+1} could be incrementalized again,
  yielding a higher-order derivative: let \m{r_1} store the result of
  \m{2x+1}, which is now an auxiliary function; then \m{r \pluseq 2x + 1}
  becomes \m{r \pluseq r_1}; to maintain \m{r_1 = 2x+1} at \m{x \pluseq 1},
  we have
\begin{equation}
  \label{eqn-x2aux}
\begin{array}{@{}l@{}}
  2(x+1)+1 = 2x+2+1 = 2x+1+2 = r_1+2\\
  \mbox{so we can do}~ r_1 \pluseq 2
\end{array}
\end{equation}
Overall, \m{r_1} is used in \m{r \pluseq r_1} and must hold the value of
\m{2x+1} for the value of \m{x} that \m{r} uses. So, one can start with
\m{r = x^2} and \m{r_1 = 2x+1} for any \m{x}, e.g., \m{x=0}, and use the
incremental maintenance
\begin{equation}
  \label{eqn-x2inc2}
\begin{array}{@{}l@{}}
  r \pluseq r_1 ~\mbox{followed by}~  r_1 \pluseq 2
\end{array}
\end{equation}
Note that there is no longer \m{2x} or \m{x+x} in the maintenance, but with
the trade-off of using more space---variable \m{r_1}.  Additionally, the
incremental maintenance (\ref{eqn-x2inc2}) no longer uses \m{x} and thus
can be done either before or after \m{x \pluseq 1}.

Note also that the entire discussion above applies for the maintenance \m{r
  \pluseq 2x-1} in (\ref{eqn-x2inc}) as well---just start with
\m{r_1=2x-1}, yielding the same \m{r \pluseq r_1}, and use
the following similar to in~(\ref{eqn-x2aux}), yielding the same
maintenance for \m{r_1}:
\[
\begin{array}{@{}l@{}}
  2(x+1)-1 = 2x+2-1 = 2x-1 +2 = r_1+2\\
  \mbox{so we can do}~ r_1 \pluseq 2
\end{array}
\]
\end{itemize}
and finally one can start with \m{r = x^2} and \m{r_1 = 2(x+1)-1=2x+1},
exactly the same initialization as above, and use the same maintenance.

While examining all the choices, consequences, and trade-offs by hand on an
ad hoc basis is challenging,
a general and systematic method can
handle these issues correctly and automatically~\cite{PaiKoe82,Liu13book}.
Because of the complexities of the matter, such a method is even more
desirable.

Note that \defn{tabulating polynomials}, used in many scientific and engineering
disciplines, is a generalization of the example of computing \m{x^2}
incrementally.  The problem is to compute the values of a polynomial
\m{f(x)} at equally spaced values of \m{x}.  Repeated incrementalization
yields exactly the same algorithm~\cite{PaiKoe82,Liu13book} as from the
\term{finite difference method}, invented by Henry Briggs in the 16th century and
used by Charles Babbage's \term{difference engine} for tabulating polynomials in
the 19th century~\cite{Goldstine:72}.
\begin{itemize}
  \setlength\itemsep{0ex}
\item The algorithm stores \m{n} variables for a polynomial of degree
  \m{n}: one for the value of the given polynomial of degree \m{n}, and one
  each for the value of a polynomial of a lower degree \m{i-1} that
  expresses the change in the value of the polynomial of degree
  \m{i}---like \m{2x+1} that expresses the change in the value of
  \m{x^2}---until degree 1.

\item The algorithm then performs \m{n} additions for a polynomial of
  degree \m{n}, generalizing the two additions in (\ref{eqn-x2inc2}) for
  \m{x^2}.
\end{itemize}
So on each incremented value of \m{x}, only \O{\m{n}} additions and no
multiplications at all are performed.

\now{\sectionmark{Optimization by incrementalization}} %
\mypar{Optimization by incrementalization---\pepm{}integration by differentiation}
\now{\sectionmark{Optimization by incrementalization}} %
\label{sec-inc-opt}

One of the big surprises at my Ph.D.\ thesis proposal examination, %
after I presented my P1 method for systematic incrementalization, was
that Anil Nerode said: ``You are doing integration by differentiation.''

I knew I was doing differentiation, and that was part of the reason I used
\m{f'} to denote the incremental version.  However, I thought integration
was the opposite of differentiation.  How could I be doing something by
doing its opposite?

It took me three years of puzzling, until I was close to finishing my
dissertation.  I had a neatly derived incremental version of the
\defn{Fibonacci function}
\[\small
fib(n) = 
\begin{array}[t]{l@{}}
  \IF n = 0 \THEN 0\\
  \IF n = 1 \THEN 1\\
  \IF n > 1 \THEN fib(n-1) + fib(n-2)
\end{array}
\]
under the input change operation of incrementing \m{n} by
1---it only adds the intermediate value of \m{fib(n-1)} to return together
with the value of \m{fib(n)} and takes \O{1} time to maintain both.
The following is how it is derived, %
with selective caching:
\begin{itemize}
\item \myemph{Identify useful intermediate results} This expands \m{fib} on
  \m{n+1} and simplifies---\m{n+1=0} to false, thus removing the first
  branch, \m{n+1=1} to \m{n=0} in the second branch,
  and so on---yielding the following, where \m{fib(n-1)} is identified as
  useful:
\[\small
fib(n+1) = 
\begin{array}[t]{l@{}}
  \IF n = 0 \THEN 1\\
  \IF n > 0 \THEN fib(n) + fib(n-1)
\end{array}
\]
\item \myemph{Extend to cache identified results} This transforms \m{fib} to
  return also the result of \m{fib(n-1)} if it is defined, yielding the
  following, where \m{fib(n) = 1st(fibExt(n))}, the first component of the
  return value:
\[\small
fibExt(n) = 
\begin{array}[t]{l@{}}
  \IF n=0 \THEN [0]\\
  \IF n=1 \THEN [1,0]\\
  \IF n>1 \LET v = fib(n-1) \IN [v+fib(n-2), \,v]
\end{array}
\]
\item \myemph{Incrementalize extended function} This defines \m{fibExt'} to
  compute \m{fibExt(n+1)}, by expanding and simplifying as for \m{fib(n+1)}
  and then replacing calls to \m{fib} using \m{fib(n)=1st(fibExt(n))} and
  \m{fib(n-1)=2nd(fibExt(n))}, the second component, yielding:
\[\small
\begin{array}{l}
fibExt'(n,rExt), \mbox{ where } rExt = fibEx(n)\\
= fibExt(n+1) = 
\begin{array}[t]{l@{}}
  \IF n=0 \THEN [1,0]\\
  \IF n>0 \LET v = 1st(rExt) \IN [v+2nd(rExt), \,v]
\end{array}
\end{array}
\]
Note that the replacements exploit both the tuple data structure and the
control structure---\m{fib(n-1)=2nd(fibExt(n))} holds for exactly \m{n>0}.
\end{itemize}

One day when I was wondering what to do with this incremental version, I
decided I could add a loop outside the incremental version to compute the
original Fibonacci function in linear time instead of the original
exponential time.  How obvious!

There it dawned on me that this is like \defn{integration by differentiation}.
That is, the integral of \m{f'} equals \m{f}, or more precisely, the
integral of \m{f'} from a base-case \m{x_0} to \m{x} plus the base-case
value \m{f(x_0)} equals \m{f(x)}.
This is a direct consequence of the %
\term{fundamental theorem of calculus}.

Most excitingly, this integration by differentiation clearly pointed to a
general method for optimization by incrementalization.  Since then,
systematic incrementalization has enabled a systematic method for program
design and optimization, discussed in Section~\ref{sec-opt}.

\mysec{Systematic design and optimization---iterate, incrementalize, implement}
\now{\chaptermark{Systematic design and optimization}}
\label{sec-opt}

\defn{Iterate-Incrementalize-Implement (III)} is a systematic method for
design and optimization. We present a vastly distilled overview of III and
how it applies to different core language features.
The method has three key steps, centered around incrementalization:
\begin{description}

\item[I1. Iterate:] determine a minimum increment to take repeatedly,
  specifically iteratively, to arrive at the desired output.

\item[I2. Incrementalize:] make expensive operations incremental in each
  iteration by using and maintaining useful additional values.

\item[I3. Implement:] design appropriate data structures for efficiently
  storing and accessing the values maintained.

\end{description}
Thanks to Tom Rothamel for picking the name III out of a combination of
choices I had in my Advanced Programming Languages course in Spring 2003.

The method was first developed for recursive functions, in the order of
Steps I2~\cite{LiuTei95Inc-SCP,LiuTei95Cir-PEPM,Liu+96Dai-POPL},
I1~\cite{LiuSto99DynProg-ESOP,LiuSto00RecToItr-PEPM}, and
I3~\cite{LiuSto02IndexDS-PEPM}. Since then, it has been used extensively in
general settings, and has proved to be drastically more powerful when used on high-level
abstractions, especially with sets and relations as high-level data abstractions.

In particular, when applied to set expressions extended with
\linebreak\term{fixed-point operations}~\cite{Paige:Henglein:87}, Steps I1, I2, and
I3 correspond to what Paige et al.\ called \term{dominated
  convergence}~\cite{Cai87thesis,CaiPai88}, \term{finite
  differencing}~\cite{Paige:Schwartz:77,Pai81thesis,PaiKoe82}, and
\term{real-time simulation}~\cite{Pai89,Cai+91,Goy00thesis}, respectively.
It is interesting to note that those were developed in the same order as
Steps I2, I1, and I3.

Table~\ref{tab-III} summarizes how the III method applies to different
language features
that provide different \term{data, control, and module
  abstractions}---loops, sets, recursion, rules, and objects---in different
programming paradigms.
\begin{table*}
  \centering
  \newcommand{\tabiiicap}{III method applied to different language features and
  abstractions.}

\pepm{}
\begin{tabular}{c|c|c}
\notnow{}
\now{Language features  & High-level    & Applying\\
                        & abstractions  & III steps}
\\\hline
loops with primitives and arrays  & none     & I2\\\hline
set expressions      & data                  & I2, I3\\\hline
recursive functions  & control               & I1, I2\\\hline
logic rules          & data, control         & I1, I2, I3\\\hline
objects with fields and methods  & module    & I2\\\hline
\end{tabular}
\arxiv{}
\now{\caption{\tabiiicap}}
\label{tab-III}
\end{table*}
In particular, Step I1 is essential when recursion as high-level {control
abstraction} is used, Step I3 is essential when sets as high-level {data
abstraction} are used, and Step I2 is essential in all cases.

Note that while loops, sets, functions, and rules can express all
computations and updates, objects are essential for building large
applications with modular components.
One can of course program with all these key language features in one
language, e.g.,~\cite{ Liu+23RuleLangInteg-TPLP}.

Details of the III method appear in~\cite{Liu13book}, with a respective
chapter for each of the main features~\notnow{}\now{\citep}[Chapters 2-6]{Liu13book}.
Incrementalization, as done in Step I2, is the core for all these features.
The analysis and transformations for incrementalization discussed in Section~\ref{sec-inc} are based
on the semantics and properties of the features used.

\mysubpar{Incrementalization is the driving core}

In fact, although it had become clear that repeated computation with an
incremental version can yield an optimized version of a given function, as
discussed in Section~\ref{sec-inc-opt}, how exactly to repeat remained
elusive for me for another three years.  That is, given a function alone,
what is the right \m{\oplus} to use?
\begin{itemize}
\item For example, for the Fibonacci function, incrementing \m{n} to
  \m{n+1} worked. How about decrementing it to \m{n-1}, or incrementing to
  \m{n+2}, or \m{n+k}, and so on?
\item For much more complicated recursive functions, including functions
  with multiple parameters, and parameters holding aggregate values like
  lists and arrays, one can see all kinds of iteration or recursion being
  used, e.g., as in dynamic programming algorithms.  Should even one or
  more parameters be considered as being changed?
\end{itemize}

This was so complicated to sort out that I started to just set things up
for simple functions of one parameter: if it is a number, use incrementing
by 1, and if it is a list, use adding an element at the head. This clearly
is not general, until one day, when I just thought about what
incrementalization does:
\begin{quote}
  Incrementalization aims to reuse.  So, now that I have a choice for
  \m{\oplus}, a \hlit{minimum} \hlit{increment} would allow maximum
  reuse. How simple!
\end{quote}
This is exactly how Step I1 came about.
Note that this is in general only a heuristic, but this simple heuristic
has worked wonders, especially for generating optimized programs from set
expressions, recursive functions, and logic rules.

Then, we need to support efficient access of the data used by incremental
maintenance.  When there is a choice for the data structures to use, we
need to design an appropriate combination based on the kinds of accesses
needed.  This is what Step I3 does.

\mypar{Loops with primitives and arrays---imperative programming}
For programs written using loops with primitives and arrays, there are no
high-level abstractions for either data or control.
\begin{itemize}

\item Loops already encode ways to iterate, and primitives and arrays
  already have direct mappings to hardware for implementation.  So there is
  little to do for Steps I1 and I3 but to adopt those.

\item Step I2 makes expensive computations on primitives and arrays in loop
  bodies incremental, using incrementalization exploiting algebraic
  properties of primitive operations and aggregate array computations.

  Note that variables holding the values of expensive computations in loops
  automatically form \term{loop invariants}.

\end{itemize}
Examples of these transformations include 
classical \term{strength reduction} that replaces multiplications with
additions~\cite{Cocke:Kennedy:77,Cooper+01} for compiler optimization, 
more general incrementalization for
primitives~\cite{Liu97Psr-IFIP,Joh+03HwDgByInc-STTT} for hardware design,
incrementalizing \term{aggregate array
computations}~\cite{LiuSto98Array-ICCL,Liu+05Array-TOPLAS} for, e.g., image
processing,
and incrementalizing more or different kinds of aggregate array
computations, e.g.,~\cite{GauRaj06,yang2021simplifying}, including
for probabilistic inference~\cite{yang2021simplifying}.

Tabulating polynomials discussed in Section~\ref{sec-inc-diff} is an
earliest classical example.

Note, however, that if the given ways to iterate and implement do not lead
to an efficient program, it is generally too difficult to find better ways
to do those, because doing so requires understanding what the loops are
computing at a higher level, but whether a loop computes a certain function
is an undecidable problem in general.
This is why we advocate higher-level data abstractions in problem
specifications when lower-level details are unnecessary.

\mypar{Set expressions---database programming}
For programs that use expressions over sets, which provide high-level data
abstractions that must be mapped to low-level data structures, Steps I2 and
I3 are essential.  The programs may still use loops or use fixed-point
operations over sets.
Note that relations as in relational databases are just sets of tuples.
\begin{itemize}

\item Fixed-point operations, if used, are first transformed into
  while-loops.  Here, Step I1 simply chooses to iterate at the minimum
  increment to a set, i.e., adding or removing one element.

\item Step I2 transforms expensive set expressions in loop bodies into
  incremental updates, using \term{auxiliary maps} to locate and retrieve
  elements as needed.

  Set expressions are so high-level that a set of rules for transforming
  particular kinds of expressions suffices for excellent
  results~\cite{PaiKoe82,Pai84}.
  For example, for set union \m{u=s\,\U\,t} under change \m{s\unioneq \{y\}},
  i.e., adding a new element \m{y} to set \m{s},
  a rule can give the incremental maintenance 
  \m{{\it if}~y \notin t: u\unioneq \{y\}}. %

  A systematic method can also derive such rules
  automatically~\cite{Liu+06ImplCRBAC-PEPM}, following P1--P3 for
  incrementalization in Section~\ref{sec-inc}.

\item Step I3 designs
  a combination of linked lists, arrays, and/or hash tables for all 
  sets,
  so that each \term{element-wise operation} on a set can be done in constant
  time. For example, for various graph traversal algorithms, this yields
  an {adjacency list representation}.

\end{itemize}
These transformations have enabled new and better algorithms to be
developed, including those by Paige et al.\ mentioned %
at the end of Section~\ref{sec-3cat} and more, %
e.g., solving regular tree grammar based
constraints~\cite{Liu+01GramSimp-SAS}, 
parametric regular path queries~\cite{%
Liu+04PRPQ-PLDI}, 
and alias analysis~\cite{Goy05},
as well as efficient implementation of tuple-pattern based
retrievals~\cite{RotLiu07Retrieval-PEPM} that translate pseudocode algorithms
into efficient C++ implementations.

The transitive closure problem in Section~\ref{sec-tc} illustrates Steps I1
to I3 for set expressions starting with the least fixed point transformed
into a while-loop.
The distributed mutual exclusion problem in Section~\ref{sec-lamutex} shows
example transformations for Step I2 for quantifications and aggregations
over sets and sequences.

Note, however, that writing appropriate fixed-point expressions, even
though they are higher-level than writing while-loops, is still
nontrivial, because 
it requires explicitly defining monotonic functions for which fixed points
can be computed by while-loops.
Writing logic rules with recursion, as discussed in
Section~\ref{sec-rules}, is even higher level.

\mypar{Recursive functions---functional programming}
For programs that use recursive functions, which provide high-level control
abstractions that must be transformed to iterations, Steps I1 and I2 are
essential.
\begin{itemize}
\item Step I1 determines a minimum increment for iteration, by selecting
  arguments of a recursive call that \term{change minimally} from the given
  function parameters, and taking the opposite of the change.

  For example, for the Fibonacci function \m{fib(n)} discussed in
  Section~\ref{sec-inc-opt} whose definition contains calls \m{fib(n-1)}
  and \m{fib(n-2)}, the minimum increment for iteration is \m{n \pluseq 1}.

\item Step I2 transforms recursive functions into incremental versions,
  which are possibly also recursive, exactly by following P1--P3 for
  incrementalization in Section~\ref{sec-inc}.

\item Step I3 selects recursive and/or indexed data structures, i.e., trees
  and/or arrays, to store results of function calls: the latter if the
  arguments of function calls can take arbitrary values, and the former
  otherwise.

\end{itemize}
These transformations have enabled systematic derivation of \term{dynamic
programming} algorithms~\cite{LiuSto99DynProg-ESOP,LiuSto03DynProg-HOSC},
more that also use arrays~\cite{LiuSto02IndexDS-PEPM},
and systematic transformation of \term{recursion to
iteration}~\cite{LiuSto00RecToItr-PEPM}.

They have interesting unexpected results on even well-known small
functions: factorial, Fibonacci, and
Ackermann~\notnow{}\now{\citep}[Chapter 4]{Liu13book}.
For example, for the Fibonacci function discussed in
Section~\ref{sec-inc-opt}, adding a loop outside the incremental
maintenance produces a well-known linear-time algorithm, but 
a better algorithm can be produced that uses one fewer variable, no copies
in the iterations, and only half the number of iterations.

Note, however, that lists and trees
as generally used in recursive functions, even though they support easy
recursive traversal of nodes in order, are often unnecessarily lower-level
and thus limiting, because they do not support easy and fast access to any
node not at the boundaries.  Writing logic rules over sets, as discussed in
Section~\ref{sec-rules}, is higher-level and overcomes this limitation.

\mypar{Logic rules---logic programming}
\label{sec-rules}
Logic rules provide both high-level data abstractions and high-level
control abstractions, because they support both sets and
recursion---predicates in rules are simply relations and thus sets of
tuples, and predicates can be defined recursively in rules.  All three
Steps I1--I3 of the III method are essential for efficient implementations
of logic rules.
\begin{itemize}

\item Step I1 first transforms rules into \term{fixed-point operations}
  over sets
  following the semantics of the rules, and then into while-loops that
  consider the minimum increment of \term{one newly inferred fact} in each
  iteration.

\item Step I2 transforms expensive set expressions in incremental updates,
  essentially as for set expressions.

\item Step I3 designs data structures for implementing set operations,
  essentially as for set expressions.

\end{itemize}
These transformations have been developed for \term{Datalog
  rules}~\cite{LiuSto03Rules-PPDP,LiuSto09Rules-TOPLAS}, supporting also
\term{on-demand
  queries}~\cite{Tek10thesis,TekLiu10RuleQuery-PPDP,TekLiu11RuleQueryBeat-SIGMOD}
and various extensions, e.g.,~\cite{TekLiu19Negation-ICLP}, especially
including precise \term{time and space complexity guarantees}.

On-demand queries use \term{demand
  transformations}~\cite{TekLiu10RuleQuery-PPDP,TekLiu11RuleQueryBeat-SIGMOD}.
They improve over \term{magic-set
  transformations} \cite{Ban+86,BeeRam91} significantly in both method
simplicity and time and space complexities.

The III method has led to new and improved algorithms or improved
complexities, e.g., for model checking~\cite{HriLiu06MCPDS-VMCAI}, secure
information flow~\cite{Hri+06InfoFlow-PLAS}, trust
management~\cite{Hri+07SPKI-PPDP}, and pointer
analysis~\cite{TekLiu16PointerAnal-TPLP}.  With demand transformations,
this has also led to many additional applications, e.g., in program
analysis, context-free grammar parsing, ontology analysis, and graph
queries in general~\cite{Tek10thesis}.

The transitive closure problem in Section~\ref{sec-tc} illustrates Steps I1
to I3 starting with Datalog rules for the problem, together with precise
complexity analysis.
The regular path query problem in Section~\ref{sec-rpq} shows decomposition
of rules that have more than two hypotheses.

Note that iterating with a minimum increment---one newly inferred fact in
each iteration---is a nontrivial improvement over what is known as
\defn{semi-naive evaluation}~\cite{bancilhon1986amateur}, which considers
all newly inferred facts in each iteration.  While semi-naive evaluation
improves over naively considering all inferred facts in each iteration,
handling the set of all newly inferred facts, instead of one such fact, in
each iteration requires additional care to avoid duplicated work within
each iteration, and maintaining the set of all newly inferred facts in each
iteration also incurs extra overhead.

\mypar{Objects with fields and methods---object-oriented programming}
\label{sec-obj}
Finally, objects provide module abstraction, hiding both data and control,
in representations of fields and implementations of methods, respectively.
Performing Step I2 across modular components is essential, because expensive
expressions may depend on data hidden in different objects.
\begin{itemize}

\item Objects encapsulate both control structures, whether low-level loops
  or high-level recursion, and data representations, whether low-level
  arrays or high-level sets. 
  Steps I1 and I3 are generally the same as already discussed, having
  little to do for
  loops and arrays and transforming within the enclosing objects for
  recursion and sets that are encapsulated within.%

\item Step I2 transforms expensive computations that use data within the
  same object as already discussed, and transforms across objects for
  expensive computations using data encapsulated and updated in different
  objects~\cite{Liu+05OptOOP-OOPSLA}.
  The latter follows the principle that each object hides its own data and
  provides methods for others to observe needed information, and gets
  notified when changes happen so as to perform incremental maintenance.

\end{itemize}
Objects help organize complex applications, and incrementalization allows
high-level queries to be used and be optimized, as discussed for a large
electronic health record system~\notnow{}\now{\citep}[Section 6.4]{Liu13book} and a robot
game~\notnow{}\now{\citep}[Section 6.5]{Liu13book}.
Incrementalization across object abstraction~\cite{Liu+05OptOOP-OOPSLA} actually yields the well-known
widely used \term{observer pattern}~\cite{GamHelJohVli95}.

For \term{complex queries over nested objects and sets}, a neat translation
into queries over sets of pairs allows incremental queries to be generated
fully automatically~\cite{Rot08thesis,RotLiu08OSQ-GPCE,Liu+16IncOQ-PPDP}:
each attribute is translated to a set of pairs, relating the object and the
attribute value, and all sets are translated to a set of pairs, relating
the object and its enclosing set.

Note that this supports that objects provide only module abstraction, not
data or control abstraction, even though they hide data and control.
Therefore, handling objects requires essentially \hlit{only a change of
  perspective}~\cite{Rot08thesis,RotLiu08OSQ-GPCE} to break through module
abstraction~\citep[Section 7.3]{Liu13book}, not separate or different
incrementalization at the core.

More importantly from a modeling perspective, interesting real-world
objects are \defn{live objects}---they are entities with not only separate
data but also separate control carrying out operations concurrently.  That
is, they are \defn{concurrent and distributed processes}, unlike objects in
traditional object-oriented programming whose operations must be run with
one or more threads, allowing nasty errors when multiple threads step on
the same data.  Incrementalization across object abstraction will ultimately
be across concurrent and distributed objects communicating through message
passing. Section~\ref{sec-lamutex} shows an example of incrementalization
for Lamport's distributed mutual exclusion algorithm.

Interestingly, studying concurrent and distributed objects has led to a
high-level language for distributed algorithms and even a simple unified
semantics for logic rules, with unrestricted negation, quantification, and
aggregation, as discussed in Section~\ref{sec-raise}.

\newenvironment{code}{\Vex{-.0}\begin{alltt}\footnotesize}{\end{alltt}\Vex{-.0}}
\newenvironment{smallcode}{\begin{alltt}\scriptsize}{\end{alltt}\vspace{-1.5ex}}

\mysec{Example applications: graph analysis and distributed algorithms}

We show example applications of systematic incrementalization and
optimization in graph analysis and distributed algorithms.  
The example problems are expressed easily at a high level, using logic
rules and high-level queries, including aggregate and quantified queries,
and queries over sent and received messages in distributed systems.
Their efficient implementations with precise complexities are derived
following the methods discussed in Sections~\ref{sec-inc}
and~\ref{sec-opt}.

We use teletype font for program code, explain the language constructs 
and transformations
informally, and give references for their precise descriptions.
We use \co{\#} at the end of a code line for comments, and \# in big-\m{O}
notation for the number of elements in a set or sequence.

\mypar{Transitive closure}
\label{sec-tc}

Given a graph with a set of edges, the problem is to compute the
\defn{transitive closure} of the graph---the set of all pairs of vertices
\co{u} and \co{v} such that there is a path from \co{u} to \co{v} following a
sequence of edges.

This is one of the most classical graph problems in algorithm textbooks.
It is used extensively in networking analysis, semantics web, program
analysis, and graph queries in general.

\mysubpar{Specification using logic rules}

Let \co{edge(u,v)} denote that there is an edge from vertex \co{u} to
vertex \co{v}, and let \co{path(u,v)} denote that there is a path from
vertex \co{u} to vertex \co{v} following a sequence of edges.
The problem can then be easily expressed as computing the set of all
\co{path} facts using two logic rules, in the simple form of what is called
{Datalog rules}:
\begin{code}
  edge(u,v) \m{\to} path(u,v)
  edge(u,w), path(w,v) \to path(u,v)
\end{code}
These rules say that, if there is an edge from vertex \co{u} to vertex
\co{v}, then there is a path from \co{u} to \co{v}; and if there is an edge
from \co{u} to \co{w}, and there is a path from \co{w} to \co{v}, then
there is a path from \co{u} to \co{v}.

The goal is to generate an efficient algorithm from the rules.  Given that
logic rules use high-level abstractions for both control and data, we need
to determine how the computation should proceed iteratively, and
incrementally, and how to represent all the data, by using the III method.

What is most interesting and important is that, because logic rules are
very high-level, we can compile the given rules into an efficient
implementation with precise time and space \hlit{complexity guarantees},
and the precise time and space complexities can be calculated just from the
rules.

\mysubpar{Step I1. Iterate---add one fact at time until a fixed point is
  reached}
The meaning of a set of rules and a set of facts is the least set of
facts that contains all the given facts and all the facts that can be
inferred using the rules.
To compute this meaning, the set of rules is first transformed into a
fixed-point computation.  Using our method, we choose the addition of
\hlit{one newly inferred fact} as the minimum increment in each iteration of
the fixed-point computation.  This generates the following while-loop:
\begin{code}
  R := \{\}                                  # initialize result set R
  while some x in e0 + e1(R) + e2(R) - R:  # while some fact x not in R
    R := R + \{x\}                           # add the new fact x to R
\end{code}
where \co{e0} is the set of given \co{edge} facts, and \co{e1} and \co{e2}
are sets of new \co{path} facts inferred using the two respective rules
given the facts in \co{R}:
\begin{code}
  e0   \=\{[edge,u,v]: edge(u,v) in givenFacts\}          # given facts
  e1(R)\=\{[path,u,v]: [edge,u,v] in R\}                  # using rule 1
  e2(R)\=\{[path,u,v]: [edge,u,w] in R, [path,w,v] in R\} # using rule 2
\end{code}
At this high level, facts are represented as tuples containing the relation
name and arguments.
\co{+} and \co{-} denote set union and difference, respectively.
\co{some x in s} returns true iff there is some element in \co{s}, and when
it returns true, it also binds \co{x} to an arbitrary element of \co{s}.

\mysubpar{Step I2. Incrementalize---use auxiliary maps indexed by shared arguments}
This transforms the while-loop to store the value of expensive
computation \co{e0 + e1(R) + e2(R) - R} in a fresh variable \co{W},
use \co{W} in place of \co{e0 + e1(R) + e2(R) - R},
initialize \co{W} to the value of \co{e0} at \co{R := \{\}}, and
incrementally update \co{W} to reflect changes to \co{e1(R) + e2(R) - R}
when an \co{edge} or \co{path} fact is added to \co{R}:
\begin{enumerate}
  \setlength\itemsep{0ex}

\item With \co{e1(R)}, for rule 1: when an \co{edge(u,v)} is added to
  \co{R}, the corresponding \co{path(u,v)} is added to \co{W} if it is not
  already in \co{R}.

\item With \co{e2(R)}, for rule 2: this needs auxiliary maps for efficient
  retrievals.
  Precisely, we define an auxiliary relation \co{edgewu} that is the
  inverse of \co{edge}, where the first component \co{w} is the shared
  argument of the two hypotheses:
\begin{code}
  edgewu\=\{[w,u]: [edge,u,w] in R\}
\end{code}

  (a) When an \co{edge(u,w)} is added to \co{R}, the matching
  \co{path(w,v)}'s are found by using \co{path} to map the \co{w} to each
  \co{v} satisfying \co{path(w,v)}; and the resulting \co{path(u,v)} is
  added to \co{W} if it is not already in \co{R}.

  (b) When a \co{path(w,v)} is added to \co{R}, the matching
  \co{edge(u,w)}'s are found by using \co{edgewu}
  to map the \co{w} to each \co{u}
  satisfying \co{edge(u,w)}; and the resulting \co{path(u,v)} is added to
  \co{W} if it is not already in \co{R}.

\end{enumerate}
This ensures that each combination of facts that makes all hypotheses of a
rule true, and thus triggers a \defn{firing} of the rule, is considered
exactly once.
Therefore, the time complexity is
the number of such combinations, or firings, summed over all rules.

For the transitive closure rules, rule 1 is fired \s{\co{edge}} of times,
but the time complexity is dominated by rule 2: 
\begin{itemize}
  \setlength\itemsep{0ex}
\item 2(a) above considers each \co{edge(u,w)} added, and finds the
  matching \co{v}'s for each, which is bounded by the number of all
  vertices, denoted \vertex;
\item 2(b) above considers each \co{path(w,v)} added, and finds the matching
  \co{u}'s for each, which is bounded by the maximum indegree of all
  vertices, denoted \indeg.
\end{itemize}
The time complexity is thus
\O{\min(\s{edge}\times\vertex,\,\s{path}\times\indeg)}.
Note that both arguments of \m{\min} are upper bounds of the number of
firings for rule 2. Note also the following:
\begin{itemize}
  \setlength\itemsep{0ex}
\item The first argument is asymptotically better than the usual algorithm
  in textbooks with a \O{\vertex^3} time, by a factor of \vertex on sparse
  graphs.
\item The second argument is bounded by the output size, and thus optimal,
  for graphs with a constant \indeg.  This output size is also achieved if
  the maximum outdegree is a constant, by simply switching \co{edge} and
  \co{path} in the conditions in rule 2.
\end{itemize}

\mysubpar{Step I3. Implement---design nested linked and indexed data structures}
This transforms operations on sets into operations on low-level linked
lists and arrays, guaranteeing that each operation takes worst-case \O{1}
time.  Note that one could simply use hash tables for sets, but that gives
only average-case \O{1} time for each operation.

For sets whose elements are atomic, one can design linked data structures %
over records, %
using what is called based representation. It creates a collection of
records, called a \defn{base}, holding elements of all sets that may
contain some same elements.  These sets are represented based on two kinds
of operations:
\begin{enumerate}
  \setlength\itemsep{0ex}

\item A set \hlit{from which elements are retrieved}, e.g., set \co{W} as
  in\linebreak \co{while some x in W}, is a linked list of pointers to the
  records that are in the set.  So retrieving each element takes \O{1}
  time.

\item A set \hlit{in which a given value must be located iff it is in the
    set}, e.g., set \co{S} as in membership test \co{x in S}, is a field in
  the records that indicates whether each record is in the set. So any such
  operation takes \O{1} time. It also makes adding or deleting any element
  to be \O{1} by locating the element first.

\end{enumerate}
These work when there is a constant number of sets with the second kind of
operations, because a record has a constant number of fields.

For sets whose elements are tuples, each value \co{v} of a component of a
tuple maps to a set of values of the next component.  Thus, the number of
such sets is not a constant.
\begin{itemize}
  \setlength\itemsep{0ex}

\item If these sets have the second kind of operations, an array \co{a} is
  used, where \co{v} is an index value of \co{a} and \co{a[v]} stores such
  a set, and such arrays may be nested if each \co{a[v]} needs to use an
  array again.

\item
  Otherwise, linked lists, possibly nested, are sufficient.

\end{itemize}
These ensure that each set operation takes worst-case \O{1} time.
Space complexity is the space for output plus the space for auxiliary maps.

Facts for different relations are first put into separate sets, for
efficient specialized representations because they generally have different
operations.  This also removes the need of the first component of the
tuples that holds the relation name.

For the transitive closure example, we first separate \co{R} into
\co{Rpath} and \co{Redge}, holding \co{path} and \co{edge} pairs that are
in \co{R}, and similarly separate \co{W} into \co{Wpath} and \co{Wedge}.
Basically, both components of \co{Rpath} and \co{Wpath} and the first
component of \co{edgeuw} use arrays, for membership test (and update) for
\co{Rpath} and \co{Wpath}, and for locating the key of maps for \co{Rpath}
and \co{edgewu}; all the rest are retrievals and use linked lists.
Precisely,
\begin{itemize}
  \setlength\itemsep{0ex}

\item
A base is used for all vertices, because all components of tuples in all
sets are those vertices.

\item 
Elements of the base are stored in an array indexed by vertices \co{1} to
\vertex, for locating
the first component of 
\co{Rpath}, \co{Wpath}, and %
\co{edgewu}. 

\item
Each element of the base array is a record of 6 fields: 2 arrays for
locating the second component of
\co{Rpath} and \co{Wpath}, %
respectively; 2 linked lists for retrieving elements in those 2 arrays;
and 2 linked lists for retrieving the second component of 
\co{Wedge} %
and %
\co{edgewu}, respectively.

\item Finally, there are 2 linked lists %
  for retrieving the first component of
\co{Wpath} and \co{Wedge}, respectively.

\end{itemize}
The output space, for \co{R}, is \O{\vertex^2}.  The auxiliary space, for
\co{edgewu} as the inverse map, is \O{\s{edge}}

Detailed transformation method, with precise resulting program and data
structures, can be found in~\cite{LiuSto09Rules-TOPLAS} and~\citep[Chapter
5]{Liu13book}.

\mypar{Regular path queries}
\label{sec-rpq}

Consider a graph with a labeled edge relation 
and a source vertex,
and a regular expression converted into a finite automata with
a labeled transition relation,
a start state,
and a final state.  A \defn{regular path query} computes all vertices
in the graph such that there is a path of labeled edges from 
the source vertex to
the target vertex
in the graph that matches a path of labeled transitions from
the start state
to a final state in the automata.

This path query problem is fundamental in both theoretical and practical
studies for database queries, program analysis, model checking, as well as
knowledge base in general.

\mysubpar{Specification using logic rules}

Let \co{edge(u,l,v)} denote that there is an edge from vertex \co{u} to
vertex \co{v} labeled \co{l}, and let \co{source} be the source vertex.
Let \co{tran(s,l,t)} denote that there is a transition from state \co{s} to
state \co{t} labeled \co{l}, and let \co{start} and \co{final} be the start
and final states, respectively.
The problem can then be easily expressed using the following rules, where
\co{match(u,s)} denotes that a path from \co{source} to \co{u} in
the graph matches a path from \co{start} to \co{s} in the automata, and
\co{return} facts are answers to the regular path query problem:
\begin{code}
  match('source','start')
  match(u,s), edge(u,l,v), tran(s,l,t) \to match(v,t)
  match(target,'final') \to return(target)
\end{code}

The goal is again to generate an efficient algorithm from the rules.  The
same method as for the transitive closure problem is used, except that this
problem has a rule with 3 hypotheses, unlike the transitive closure
problem, which has rules with only 1 or 2 hypotheses.

\mysubpar{Decomposing rules with more than two hypotheses}

For any rule with more than 2 hypotheses, it can be decomposed into rules
with 2 hypotheses each by introducing new relations as intermediate or
auxiliary values.  Simply take 2 hypotheses and conclude a new relation,
with arguments including variables used in the remaining hypotheses, i.e.,
omitting variables used in only the 2 hypotheses taken.

For regular path queries, the second rule can be decomposed as follows, by
simply considering the first 2 hypotheses first:
\begin{code}
  match(u,s), edge(u,l,v) \to matchedge(s,l,v)
  matchedge(s,l,v), tran(s,l,t) \to match(v,t)
\end{code}
where \co{matchedge} is a new relation, and variable \co{u} is omitted from
its arguments.

There are 2 other ways of decomposition, by considering the last 2
hypotheses first, yielding:
\begin{code}
  edge(u,l,v), tran(s,l,t) \to edgetran(u,v,s,t)
  match(u,s), edgetran(u,v,s,t) \to match(v,t)
\end{code}
and by consider the first and last hypotheses first, yielding:
\begin{code}
  match(u,s), tran(s,l,t) \to matchtran(u,l,t)
  matchtran(u,l,t), edge(u,l,v) \to match(v,t)
\end{code}

Which decomposition to use depends on precise complexity analysis of the
generated implementation based on just the rules, as shown next.

\mysubpar{III steps and resulting time and space complexities}

The III transformations are similar as for the transitive closure problem.
The only different is, for rules with 2 hypotheses, auxiliary relations
are used to map from one or more variables shared between the 2
hypotheses to the remaining one or more variables, not limited to only one
variable as for the transitive closure problem.

The complexities are dominated by the 2 rules decomposed from the second
rule. Let \state be the number of states in the automata, \vertex be the
number of vertices in the graph, and \lbl be the number of labels in both.

Consider the first set of 2 rules.  The number of firings of rule 1 is
bounded by \s{edge} multiplied by the number of matching values for variable
\co{s}, and that of rule 2 is bounded by \s{tran} multiplied by the number
of matching values for variable \co{v}.  Thus, the worst-case time
complexity is bounded by
\O{\s{edge} \times \state + \s{tran} \times \vertex}.
The last set of 2 rules can be analyzed in a similar way, and its
worst-case time complexity has exactly the same bound.

Consider the second set of 2 rules.  The newly introduced relation
\co{edgetran} from rule 1 is exactly the well-known product graph of
\co{edge} and \co{tran}, and rule 2 is exactly the well-known reachability
rule in the product graph.  The number of firings of rule 1 dominates,
because rule 2 only filters the result \co{edgetran} from rule 1.  The
worst-case time complexity is the minimum of \s{edge} multiplied by all
matching transitions and \s{tran} multiplied by all matching edges, both
bounded by
\O{\s{edge} \times \s{tran}}.

The space complexity for inferred \co{match} and \co{return} is the same
for all 3 sets of rules, and is dominated by \co{match}, bounded by
\O{\vertex\times\state}.
The main overall dominate space is the different space for the introduced
new relations: \co{matchedge} and \co{matchtrans} for the first and third
sets of rules are both bounded by \O{\lbl\times\vertex\times\state}, and
the product graph \co{edgetrans} for the second set of rules is bounded by
\O{\vertex^2\times\state^2}.

While there are trade-offs in general, \lbl is typically small and each
label may connect a nontrivial number of pairs of vertices and pairs of
transitions.
Therefore, the time complexity for the first and third decompositions is
asymptotically better when the edges and transitions are dense.  The space
complexity for them is always asymptotically much better, having avoided
building the product graph.

As a fourth way, one could also consider all 3 hypotheses of the second rule
of together without decomposition, by using the method
in~\cite{Liu+06ImplCRBAC-PEPM}.  This yields an algorithm with worst-case
still bounded by \O{\s{edge} \times \s{tran}} as in the second set of 2
rules above, but with better space complexity by not storing a new
relation.  Analysis that compares the complexities of all 4 ways can be
found in~\cite{LiuSto09Rules-TOPLAS}.

\mypar{Distributed mutual exclusion}
\label{sec-lamutex}

In a distributed system where a set of processes need to access a shared
resource, \defn{distributed mutual exclusion} ensures that the processes
access the shared source mutually exclusively, in what is called a
critical section, i.e., there can be at most one process in a critical
section at a time.  We consider Lamport's algorithm for distributed mutual
exclusion, which was developed to illustrate the logical clock he invented
~\cite{Lam78}.

\mysubpar{High-level description in English}

Figure~\ref{fig-lam-paper} contains Lamport's original description of the
algorithm, except with the notation \m{<} instead of \m{\Rightarrow} in
rule~5 (for comparing pairs of logical time and process id using lexical
ordering: %
\co{(t,p)\,\m{<}\,(t2,p2)} iff \co{t\,\m{<}\,t2} or \co{t\,\m{=}\,t2} and
\co{p\,\m{<}\,p2}%
) and with the word ``acknowledgment'' added in rule~5 (for simplicity when
omitting a commonly omitted~\cite{Lynch96,Garg02} small optimization
mentioned in a footnote).%
\footnote{Lamport's original description has a separate paragraph describing
initializations but it is more complex than necessary, and 
initializing the queue in each process to empty suffices.}

\begin{figure}[htbp]
  \centering
  \footnotesize
\fbox{
\begin{tabular}{@{}p{0.95\textwidth}@{}}

\quad The algorithm is then defined by the following five rules.  For
convenience, the actions defined by each rule are assumed to form a
single event.

\quad 1. To request the resource, process \m{P_i} sends the message
{\it \m{T_m}:\m{P_i} requests resource} to every other process, and puts that
message on its request queue, where \m{T_m} is the timestamp of the
message.

\quad 2. When process \m{P_j} receives the message {\it\m{T_m}:\m{P_i} requests
  resource}, it places it on its request queue and sends a (timestamped)
acknowledgment message to \m{P_i}.

\quad 3. To release the resource, process \m{P_i} removes any
{\it\m{T_m}:\m{P_i} requests resource} message from its request queue and
sends a (timestamped) {\it\m{P_i} releases resource} message to every
other process.

\quad 4. When process \m{P_j} receives a {\it\m{P_i} releases resource}
message, it removes any {\it\m{T_m}:\m{P_i} requests resource} message from
its request queue.

\quad 5. Process \m{P_i} is granted the resource when the following two
conditions are satisfied: (i) There is a {\it\m{T_m}:\m{P_i} requests
  resource} message in its request queue which is ordered before any
other request in its queue by the relation \m{<}. (To define the
relation \m{<} for messages, we identify a message with the event of
sending it.) (ii) \m{P_i} has received an acknowledgment
message from every other process timestamped later than \m{T_m}.

Note that conditions (i) and (ii) of rule 5 are tested locally by
\m{P_i}.
\end{tabular}
}
  \caption{Lamport's description in English.}
  \label{fig-lam-paper}
\end{figure}

The algorithm assumes FIFO and reliable channels for message passing. The
algorithm
is safe in that at most one process can be in a critical section at a time.
It is live in that some process will be in a critical section if there are
requests.  It is fair in that requests are granted in the order of \m{<},
on pairs of logical time and process id, of the requests.
Its message complexity is \m{3(n-1)} in that \m{3(n-1)}
messages are required to serve each request, where \m{n} is the number of
processes.

\mysubpar{Specification using high-level queries} %

Figure~\ref{fig-lam-orig} shows Lamport's algorithm expressed in a simple
way in DistAlgo~\cite{Liu+17DistPL-TOPLAS}, a high-level language for
distributed algorithms.\footnote{A few lines for starting the processes,
  and for running some critical-section tasks, could be added to make a
  complete program~\cite{Liu+17DistPL-TOPLAS}, but they are irrelevant to
  the optimizations.}
The %
description in Figure~\ref{fig-lam-paper} corresponds to the
body of \co{mutex} and the two \co{receive} definitions in
Figure~\ref{fig-lam-orig}. %

\begin{figure}[htbp]
\begin{smallcode}
 1 class P extends process:           # define process type P
 2   def setup(s):                    # take set s of all other processes
 3     self.q := \{\}                   # initialize set q of pending requests

 4   def mutex(task):                 # run task with mutual exclusion
 5     -- request
 6     self.t := logical_time()                                                # rule 1
 7     send ('request', t, self) to s                                          #
 8     q.add(('request', t, self))                                             #
 9     await each ('request',t2,p2) in q | (t2,p2) != (t,self) implies (t,self) < (t2,p2)
10           and each p2 in s | some received('ack', t2, =p2) | t2 > t         # rule 5
11     task()                         # critical section task
12     -- release
13     q.del(('request', t, self))                                             # rule 3 
14     send ('release', t, self) to s                                          #  

15   receive ('request', t2, p2):                                              # rule 2
16     q.add(('request', t2, p2))                                              #
17     send ('ack', logical_time(), self) to p2                                #

18   receive ('release', t2, p2):                                              # rule 4
19     q.del(('request', t2, p2))                                              #
\end{smallcode}
  \caption{Lamport's algorithm (lines 5--19) in DistAlgo.}
  \label{fig-lam-orig}
\end{figure}

It is easy to see that all
operations %
in Figure~\ref{fig-lam-orig} are simple and inexpensive except for the two
conditions in \co{await} for rule 5, on lines 9--10.
These conditions use high-level queries over sets and sequences, including
message histories.  In particular,
\begin{itemize}
  \setlength\itemsep{0ex}
\item
\co{each x in s | cond(x)} and
\co{some x in s | cond(x)} are universal
and existential quantifications, respectively.
\item
\co{received} is the sequence of all messages received, and
\co{received(m)} denotes \co{m in received}.
\item
\co{=x} in a tuple is an equality test against \co{x}, e.g.,
\co{some (x,=y) in s |} \co{cond(x,y)} denotes
\co{some (x,y2) in s | y2=y and cond(x,y)}.
\end{itemize}

\mysubpar{Incrementalize expensive queries}
Consider the two expensive conditions in \co{await} in
Figure~\ref{fig-lam-orig}, and transform the first one to have an equivalent
simplified condition inside:
\begin{code}
  each ('request',t2,p2) in q | (t,self) <= (t2,p2)
\end{code}
and
\begin{code}
  each p2 in s | some received('ack',t2,=p2) | t2 > t
\end{code}
They take \O{\s{q}} and \O{\s{s}\times\s{received}} time, respectively.  To
optimize, we incrementalize them when the variables they use are updated.

We first \hlit{identify all updates} to variables used by these expensive
queries, yielding:
\begin{code}
  self.s = s                     # automatic when s is passed in setup
  self.q = \{\}                    # line 3
  self.t = logical_time()        # line 6
  q.add((’request’, t, self))    # line 8
  q.del((’request’, t, self))    # line 13
  q.add((’request’, t2, p2))     # line 16
  q.del((’request’, t2, p2))     # line 19
  received.add((’ack’,t2,p2))    # automatic when receiving an ack msg
\end{code}

We then 
\hlit{transform quantifications into aggregations} \co{count}, \co{min},
\co{sum}, etc.
Each form of quantified queries may have a few ways to transform into
aggregate queries using different aggregate operations.
Which form to use depends on the costs and frequencies of the optimized
query and incremental maintenance under the updates.

For example, the two conditions above can be transformed to use \co{min}
and \co{count}, respectively:
\begin{code}
                  \{(t2,p2): ('request',t2,p2) in q\} == \{\} or
  (t,self) <= min \{(t2,p2): ('request',t2,p2) in q\}
\end{code}
and
\begin{code}
  count \{p2: p2 in s, (’ack’,t2,=p2) in received, t2 > t\} == count s
\end{code}\Vex{-3}
\begin{itemize}
  \setlength\itemsep{0ex}

\item 
Maintaining \co{min} under updates to \co{q} can use a queue data
structure, say called \co{ds}, that handles each addition and deletion in
\O{\log \s{q}} time and returns \co{min} in \O{1} time while holding the set
\begin{code}
  \{(t2,p2): ('request',t2,p2) in q\}
\end{code} 
\item
Maintaining the two \co{count}'s under updates to \co{s}, \co{t}, and
\co{received} uses and incrementally maintains the following values:
\begin{code}
  acked = \{p2: p2 in s, (’ack’,t2,=p2) in received, t2 > t\}
  count_ack = count acked
  total = count s
\end{code}
by setting \co{total} when \co{s} is passed in; setting \co{acked := \{\}}
and \co{count\_ack := 0} when a new time \co{t} is taken, which is larger than
all previous times; and updating \co{acked} and \co{count\_ack} when
a new \co{ack} message is added to \co{received}, all in \O{1} time.
\end{itemize}
The two conditions become the following, taking only \O{1} time:
\begin{code}
  (ds.is_empty() or (t,self) <= ds.min()) and count_ack == total
\end{code}
Figure~\ref{fig-lam-inc-min} shows the resulting optimized program.  It is
the same as Figure~\ref{fig-lam-orig} except that the \co{await} conditions
are optimized, all other lines with comments are added, and all lines with
\co{q} are removed as \co{q} is no longer used.

\begin{figure}[htbp]
\begin{smallcode}
 1 class P extends process:
 2   def setup(s):
 3     self.total := count s                          # total num of other processes
 4     self.ds := new DS()                            # data structure for maintaining
                                                      #   all requests
 5   def mutex(task):
 6     -- request
 7     self.t := logical_time()
 8     self.acked := \{\}                               # set of acked processes
 9     self.count_ack := 0                            # num of acked processes
10     send ('request', t, self) to s
11     ds.add((t2,p2))                                # add to data structure
12     await (ds.is_empty() or (t,self) <= ds.min())  # use maintained results
13           and count_ack = total                    # use maintained results
14     task()
15     -- release
16     ds.del((t2,p2))                                # delete from data structure
17     send ('release', logical_time(), self) to s

18   receive ('request', t2, p2):
19     ds.add((t2,p2))                                # add to data structure
20     send ('ack', logical_time(), self) to p2

21   receive ('ack', t2, p2):                         # new message handler
22     if t2 > t:                                     # comparison in conjunct 2
23       if p2 in s:                                  # membership in conjunct 2
24         if p2 not in acked:                        # if not acked already
25           acked.add(p2)                            # add to acked
26           count_ack +:= 1                          # increment count_ack

27   receive ('release', t2, p2):
28     ds.del((t2,p2))                                # delete from data structure
\end{smallcode}
\caption{Optimized program with use of \co{min} after incrementalization.}
  \label{fig-lam-inc-min}
\end{figure}

This optimized program matches the intended use of queue in the original
description in English, and both the \hlit{time and space are drastically
  improved}.  Two conditions are optimized from \O{\s{q}} and
\O{\s{s}\times\s{received}}, respectively, to \O{1} time, with each update
to \co{q} taking slightly increased time from \O{1} to \O{\log \s{q}} for
incremental maintenance.  Overall this is a drastic improvement, because
for each request, the two conditions are tested repeatedly and the updates
are executed once for addition and once for deletion,
and furthermore, \co{received} would
grow unboundedly but is removed in the optimized program as it is no longer
used.

Nevertheless, interestingly, a \hlit{better optimized program} can be
derived.  There is another way of transforming the first condition to
aggregation that uses \co{count}, not \co{min}:
\begin{code}
  count \{ (’request’,t2,p2) in q: (t,self) > (t2,p2) \} == 0
\end{code}
Incrementally maintaining this query result can be done similarly as for
maintaining the \co{count\_ack} above, in \O{1} time, without using any queue
that takes \O{\log \s{q}} time.

Detail of the transformation method for quantifications and the complete
better optimized program can be found
in~\cite{Liu+12DistPL-OOPSLA,Liu+17DistPL-TOPLAS}.
One can also derive a slight variant of the optimized program that uses
queue but does not put a process's own request in the
queue~\cite{Liu+17DistPL-TOPLAS}.
Additionally, un-incrementalization of the original program even leads to a
simplified original program without \co{q} at all, by more uses of
\co{received} inside
\co{await}~\cite{Liu+12DistPL-OOPSLA,Liu+17DistPL-TOPLAS}.

\mysec{Conclusion---raising the level of abstractions}
\label{sec-raise}

We have discussed that incrementalization is the essence of incremental
computation and, even more importantly, is the core of a systematic method for
program design and optimization, including algorithm design and
optimization.

\mypar{High-level abstractions}
The key meta-level observation is that high-level abstractions enable
drastically easier and more powerful optimization.  In particular:
\begin{itemize}

\item High-level data abstractions using sets, including relations and
  predicates, enable high-level declarative queries, leading to not only
  efficient incremental maintenance of sophisticated database views but
  also new and better algorithms for challenging problems in complex
  applications.

  They also obviate unnecessary uses of error-prone loops and tedious
  recursion over low-level data.

\item High-level control abstractions using recursion, especially when used
  with high-level data abstraction, allow sophisticated analysis and
  queries over complex transitive relations to be expressed clearly, and be
  implemented efficiently using incrementalization for fixed-point
  computations. %

  They furthermore enable automatic calculation of precise complexity
  guarantees that is drastically harder, if possible, otherwise.

\item High-level module abstractions using objects, by encapsulating both
  data and control so that all queries and updates on the data are
  organized together, help make the analysis and transformations more
  localized.

  They also force objects that interact with each other to follow
  established patterns for general extensibility.

\end{itemize}

\mysubpar{A language for distributed algorithms}
Most interestingly, because objects can be concurrent and distributed,
studying distributed algorithms has led to the creation of a powerful
language, DistAlgo, for expressing distributed algorithms at a high
level~\cite{Liu+12DistPL-OOPSLA,Liu+17DistPL-TOPLAS}, implemented by
extending the Python programming language and
compiler.\footnote{https://github.com/DistAlgo/distalgo}
This is the language we used in Section~\ref{sec-lamutex}.

In particular, expressing distributed algorithms using high-level queries
of message histories, i.e., sent and received messages, reveals a
substantial need of logic quantifications.
Systematic incrementalization and optimization for
quantifications~\cite{Liu+12DistPL-OOPSLA,Liu+17DistPL-TOPLAS}, as we
illustrated in Section~\ref{sec-lamutex}, helped enable the support of
high-level queries in DistAlgo.

Using high-level queries and systematic incrementalization has led to
improved distributed algorithm specifications that have further led to
improved algorithms, in both safety and liveness, for many algorithms,
e.g.,~\cite{Liu+12DistSpec-SSS,Liu+17DistPL-TOPLAS,Liu+19Paxos-PPDP,shi+23DerechoDA-ApPLIED}.

\mysubpar{A unified semantics for logic rules}
Most unexpectedly, efficient implementations of quantifications have further led
to the discovery and development of a unified semantics for logic rules
with unrestricted negation, quantification, and
aggregation~\cite{LiuSto20Founded-JLC,LiuSto22RuleAgg-JLC}, including
knowledge units for using the semantics at
scale~\cite{LiuSto21LogicalConstraints-JLC}.

The new semantics not only unifies disagreeing previous semantics but also
is much simpler and exponentially more expressive.
For example, the most well-known challenging company control
problem~\cite{Ceri:Gottlob:Tanca:90,ross1997monotonic,faber2011semantics,gelfond2019vicious,colombo2025template}
is merely a straightforward least fixed point.
Interesting game problems that generalize the well-known win-not-win
game~\cite{LiuSto18Founded-LFCS,LiuSto20Founded-JLC} also require only
least fixed-point computations, but other semantics and systems that we
tried all failed to solve these problems correctly.

Experiments with examples implemented with optimization by
incrementalization also show superior performance over the best well-known
systems when they could compute correct answers, e.g., for the company
control problem, and that on some problems, e.g., the generalized
win-not-win games, none of those systems could compute correct answers, due
to implementation bugs as well as fundamentally incorrect semantics and
inference~\cite{LiuSto22RuleAgg-JLC}.

Alda~\cite{ Liu+23RuleLangInteg-TPLP}, a language extending DistAlgo with
logic rules,\footnote{https://github.com/DistAlgo/alda} was developed with
a design that supports the unified semantics.

\mypar{Limitation and future work}

Of course, systematic incrementalization will not be able to derive all most
efficient
algorithms, because whether a program can be made more efficient
is in general an undecidable problem.  However, by
making the design systematic and automated for everything that can be
automated, designers and developers can focus on truly creative work. 

Overall, tremendous effort is needed to support high-level abstractions in
widely-used languages, and implement powerful analysis and transformations
in wildely-used compilers.
The goal is to support rapid development of algorithms and generation of
programs with both correctness and efficiency guarantees.

Additional technical questions include: Can we raise the level of abstraction
even higher and generate even better algorithms and programs in better
ways?
In particular, can we program with higher-level constraints and derive
efficient programs that find desired solutions with complexity guarantees?
Also, can distributed algorithms and programs be derived systematically
from desired global properties?

Practical implementation questions include: How can we maintain compiler
extensions and optimizations when program representations in the compiler
keep changing? %
Can we make practical compiler construction for rich languages much easier,
supporting logic rules for analysis, and transformation rules for
optimizations, in a powerful language?

Ultimately, language abstractions should be at the level of natural
languages but with rigorous semantics, to allow both most effective uses
of and most critical advances in large language models
(LLMs)~\cite{Liu24RLM-LPOP}.
Powerful and rigorous transformation systems along the lines of
APTS~\cite{Paige:83,Paige94view}, KIDS~\cite{Smith:90,Smith:91},
CACHET~\cite{Liu95CACHET-KBSE}, and
InvTS~\cite{Liu+09Inv-GPCE,Gor+12Compose-PEPM} are highly desired for
generating efficient implementations.

\notnow{}
\now{
\begin{acknowledgements}
}
I am deeply grateful for all the advice and comments by many colleagues on
our work on incrementalization, %
but especially
Anil Nerode, for amazing insight and guidance for over 30 years;
Cordell Green, 
Neil Jones,
Jack Schwartz, 
and Michel Sintzoff, for great encouragement for my book~\cite{Liu13book};
Jon Barwise,
Olivier Danvy, %
Robert Dewar, 
Fritz Henglein,
and Moshe Vardi, for special encouragement for our work;
Bob Paige, 
Tom Reps,
Doug Smith, 
and Tim Teitelbaum, for excellent related work and advice;
and Scott Stoller, for all the help and collaboration for over 30 years.
Many thanks to new brilliant work by all my Ph.D.\ students, especially 
Tom Rothamel,
Tuncay Tekle,
and Bo Lin,
for amazingly neat incremental object and set queries,
demand-driven Datalog queries, and
DistAlgo compiler and optimization, respectively, discussed in this writing.
Special thanks to Paul McJones for creating critical historical software
archives at the Computer History Museum.
Additional thanks to Fritz Henglein, Serene Leona, Tom Rothamel, Scott
Stoller, Tim Teitelbaum, Yi Tong, and anonymous reviewers of Foundations
and Trends in Programming Languages, for helpful and detailed comments on
drafts of this writing.

The work of my group has been funded by continuous grants from National
Science Foundation, Office of Naval Research, and industry.
The work of this writing
was supported in part by NSF under grant CCF-1954837. %
The first version~\cite{Liu24IncEssence-PEPM} was done in part while 
I was
visiting the Simons Institute for the Theory of Computing in Fall 2023.
Thanks to the Simons Institute for the Theory of Computing and organizers
of the program on Logic and Algorithms in Database Theory and AI for
bringing wonderful diverse communities together.
\now{
\end{acknowledgements}
}

\arxiv{}

\notnow{} %

\now{
\backmatter 
\printbibliography
}

\end{document}